\documentclass[a4paper]{report}
\usepackage[utf8]{inputenc}
\usepackage[T1]{fontenc}
\usepackage{RJournal}
\usepackage{amsmath,amssymb,array}
\usepackage{booktabs}

\usepackage{algorithm2e}
\usepackage{float}
\RestyleAlgo{ruled}
\usepackage{titlesec}
\usepackage{xcolor}
\usepackage{multirow}
\renewcommand{\thesection}{\arabic{section}}
\def\thesubsection{\arabic{section}.\arabic{subsection}}
\titleformat{\section}[hang]{\Large}{\thesection\ }{2pt}{}
\titleformat{\subsection}[hang]{\large}{\thesubsection\ }{2pt}{}

\begin{document}

%% do not edit, for illustration only
\sectionhead{Article}
\volume{}
\volnumber{}
\year{}
\month{}

%% replace RJtemplate with your article
\begin{article}
% !TeX root = RJwrapper.tex
\title{SNSeg: An R Package for  Time Series Segmentation via Self-Normalization}
\author{Shubo Sun, Zifeng Zhao, Feiyu Jiang, Xiaofeng Shao}

\maketitle

\abstract{
	Time series segmentation aims to identify potential change-points in a sequence of temporally dependent data, so that the original sequence can be partitioned into several homogeneous subsequences. It is useful for modeling and predicting non-stationary time series and is widely applied in natural  and social sciences. Existing segmentation methods primarily focus on only one type of parameter changes such as mean and variance, and they typically depend on laborious tuning or smoothing parameters, which can be challenging to choose in practice. The self-normalization based change-point estimation framework SNCP by \cite{zhao2021segmenting}, however, offers users more flexibility and convenience as it  allows for change-point estimation of different types of parameters (e.g. mean, variance, quantile and autocovariance) in a unified fashion, and requires effortless tuning.  In this paper, the R package \CRANpkg{SNSeg} is introduced  to implement  SNCP for segmentation of univariate and multivariate time series.  An extension of SNCP, named SNHD, is also designed and implemented for change-point estimation in the mean vector of high-dimensional time series. The estimated change-points as well as segmented time series are available with graphical tools.  Detailed examples of \pkg{SNSeg} are given in simulations of multivariate autoregressive processes with change-points.
}

\section[Intro]{Introduction}\label{sec:intro}
Time series segmentation, also known as change-point estimation in time series, has become increasingly popular in various fields such as statistics, bioinformatics, climate science, economics, finance, signal processing, epidemiology, among many others. As a result, numerous methods have been proposed to address different types of change-point estimation problems under various settings. This in turn  leads to the development of many \textbf{R} packages for their implementation. 

{
	Here, we list some commonly used and influential packages for change-point analysis in the \textbf{R} programming language. The package \pkg{strucchange} \citep{zeileis2002struc} employs algorithms proposed by \cite{zeileis2003cmpstat} to identify structural changes in linear regression models. The package \pkg{changepoint} \citep{changepoint} provides numerous methods for estimating change-points in a univariate time series, containing the Binary Segmentation (BS) and the pruned exact linear time (PELT) algorithm as described in \cite{Killick2012}, and the segment neighbourhoods algorithm in  \cite{auger1989algorithms}. The package \pkg{mosum} \citep{mosum2021} executes the moving sum (MOSUM) procedure introduced by \cite{eichinger2018mosum} for univariate time series. It can implement MOSUM with a single bandwidth parameter and also allows multiple bandwidths via either bottom-up merging or localized pruning. The package \pkg{cpss} \citep{wang2022cpss} focuses on change-point estimation in various generalized linear models utilizing the sample-split strategy proposed by \cite{zou2020consistent}. We note that there are also packages targeting nonparametric distributional changes, e.g.  the package \pkg{ecp} \citep{ecp2014} and \pkg{cpm} \citep{cpm2015}.
}

	However, the aforementioned methods, as well as their implementation packages, are subject to certain limitations when applied to change-point estimation in multivariate time series. First, most packages only provide functions to detect specific types of changes (e.g. mean or variance). { Although it may be possible to modify these functions to cover other types of changes (e.g. quantiles), such a generalization is usually not easy and requires non-trivial effort.} This means that for the same dataset, different methods { or substantial modifications to the existing codes} may be required for estimating different types of changes, which may incur inconvenience of implementation for practitioners.  Second, { many packages implement methods that assume temporal independence among data, which may not be realistic in practice, and thus may suffer from issues such as false positive detections.}
	
	%  and may not be easily generalized or implemented when changes in different aspects of data are of interest
	
	%When dealing with time series data, these methods require laborious tuning on smoothing parameters, such as the bandwidth for consistent estimation of the long-run variance.} How to choose these tuning parameters is important yet challenging in practice.

Recently, \cite{zhao2021segmenting} have developed a new framework called  self-normalization  based change-point estimation (SNCP) to overcome the above limitations. The most appealing feature of SNCP is its versatility as it allows for change-point estimation in a broad class of parameters
(such as mean, variance, correlation, quantile and their combinations) in a unified fashion.  The basic idea of SNCP is to augment the conventional cumulative sum  (CUSUM) statistics with the technique called self-normalization (SN). SN is originally introduced by \cite{shao2010self} for confidence interval construction of a general parameter in the stationarity time series setting, and is later extended to change-point testing by \cite{shaozhang2010testing}. It can bypass the issue of bandwidth selection in the consistent long-run variance estimation. See \cite{shao2015} for a review. SNCP is fully nonparametric, robust to temporal dependence and applicable universally for various parameters of interest for a multivariate time series. Furthermore, based on  a series of carefully designed nested local-windows, SNCP can isolate each true change-point adaptively and achieves the goal of multiple change-point estimation with respectable detection power and estimation accuracy. 

In this paper, we  introduce the \textbf{R} package \CRANpkg{SNSeg} \citep{SNSeg2023}, which implements the SNCP framework in \cite{zhao2021segmenting} for univariate and multivariate time series segmentation. This is achieved by the functions \texttt{SNSeg\_Uni()} and \texttt{SNSeg\_Multi()}, respectively. Another contribution of this paper is to extend the SNCP framework to change-point estimation in the mean vector of a high-dimensional time series. Since SNCP is only applicable to fixed-dimensional time series, a new procedure  based on U-statistics \citep{wang2022inference}, termed as SNHD, is proposed by modifying the original SNCP in \cite{zhao2021segmenting}. The implementation of SNHD is available through the function \texttt{SNSeg\_HD()}. Graphical options are also allowed for plotting the estimated change-points and associated test statistics. 

The rest of the paper is organized as follows. We first provide the background of SN based statistics and the SNCP/SNHD procedures for change-point estimation in Section \ref{sec:SNCP}. In Section \ref{sec:SNSeg}, we demonstrate the core functions of the package \pkg{SNSeg} by various examples of change-point estimation problems. {Additional simulation results and comparison with existing packages are provided in Section \ref{sec:additional}.} Section \ref{sec:conclusion} concludes.

\section[SNCP Framework]{SNCP  Framework}\label{sec:SNCP}
This section gives necessary statistical backgrounds of SNCP in change-point estimation problems. We first demonstrate how an SN based CUSUM test statistic works for estimating  a single change-point, and then introduce the  nested local-window based SNCP algorithm for multiple change-point estimation.  The extension of SNCP to change-point estimation in high-dimensional mean problem is also provided and we term the related algorithm as SNHD. The issue of how to choose tuning parameters is also discussed.  

\subsection{Single Change-Point Estimation}\label{subsec:SNCPsingle}
Let $\{Y_t\}^n_{t=1}$ be a sequence of multivariate time series of dimension $p$, which is assumed to be fixed for now.
We aim to detect whether there is a change-point in the quantities $\{\theta_t\}^n_{t=1}$ defined by $\theta_t = \theta(F_t)\in \mathbb{R}^d$, where $F_t$ denotes the distribution function of $Y_t$, and $\theta(\cdot)$ is a general functional such as    mean, variance, auto-covariance, quantiles, etc.   More specifically, if there is no change-point, then \begin{flalign}\label{hypothesis_0}\theta_1=\cdots=\theta_n.\end{flalign} Otherwise, we assume there is an unknown change-point $k^*\in \{1,\cdots,n-1\}$  defined by 
\begin{flalign}\label{hypothesis_a}
	\theta_1=\cdots=\theta_{k^*}\neq \theta_{k^*+1}=\cdots=\theta_n,
\end{flalign}
and our interest is to recover the location $k^*$.

The above setting allows for at most one change-point in
$\{\theta_t\}^n_{t=1}$. A commonly used statistic for testing the existence of change-points  is based on the CUSUM process, defined by 
\begin{equation}\label{cusum}
	D_n(k)=\frac{k (n-k)}{n^{3/2}}\left(\hat{\theta}_{1,k}-\hat{\theta}_{k+1,n}\right), ~~ k\in \{1,2,\cdots,n-1\},
\end{equation}
where for any $1\leq a<b\leq n$,  $\hat{\theta}_{a,b}=\theta(\hat{F}_{a,b})$ estimates the model parameter with $\hat{F}_{a,b}$ being the empirical distribution of $\{Y_t\}_{t=a}^b$.  For example, when $\theta(\cdot)$ is the mean functional,  it can be shown that 
$$
D_n(k)=\frac{1}{\sqrt{n}}\sum_{t=1}^{ k}(Y_t-\Bar{Y}), \quad \Bar{Y}=n^{-1}\sum_{t=1}^n Y_t.
$$

The CUSUM process  in \eqref{cusum} sequentially compares the estimates before and after a time point $k$, and its norm is expected to attain the maximum when $k=k^*$. Intuitively, if \eqref{hypothesis_0} holds, then $D_n(k)$ should fluctuate around zero; otherwise if \eqref{hypothesis_a} holds, then $\hat{\theta}_{1,k}$ and $\hat{\theta}_{k+1,n}$ are consistent estimators for $\theta_1$ and $\theta_n$, respectively at $k=k^*$, and the resulting contrast $\|D_n(k^*)\|$ would be most informative about the change signal $\Delta_n=\theta_{k^*+1}-\theta_{k^*}$. Therefore, it is natural to estimate the change-point location via 
\begin{equation}\label{cusum_k}
	\tilde{k}=\arg\max_{k=1,\cdots,n-1}\|D_n(k)\|^2.
\end{equation}
However,   analyzing the  asymptotic 
distribution of CUSUM process $\{D_n(\lfloor n r\rfloor)\}_{r\in[0,1]}$ for time series data is  difficult, as it typically depends on a nuisance parameter called long-run variance \citep{newey1986simple,andrews1991heteroskedasticity}. As mentioned before, the estimation of long-run variance is quite challenging even in a stationary time series, let alone the scenario when a change-point is present.

To bypass this issue, \cite{zhao2021segmenting} propose to estimate the change-point via the self-normalized version of \eqref{cusum_k}, i.e. 
\begin{equation}\label{sn_k}
	\hat{k} = \arg \max_{k=1,\cdots,n-1}T_n(k),\quad T_n(k)=D_n(k)'V^{-1}_n(k)D_n(k),
\end{equation}
where \begin{equation}\label{V}
	V_n(k) =\sum_{i=1}^{k}\frac{i^2(k-i)^2}{n^2 k^2}(\hat{\theta}_{1,i}-\hat{\theta}_{i+1,k})^{\otimes 2} + \sum_{i=k+1}^{n}\frac{(n-i+1)^2(i-k-1)^2}{n^2(n-k)^2}(\hat{\theta}_{i,n}-\hat{\theta}_{k+1,i-1})^{\otimes 2},
\end{equation} is defined as the self-normalizer of $D_n(k)$ with $a^{\otimes 2}=aa^\top$ for a vector $a.$ The self-normalizer $V_n(k)$ is proportional to long run variance, which gets canceled out in the limiting null distribution of $T_n(k)$. Thus, the testing/estimation of a single change-point is completely free of tuning parameters.

In practice, we may not know whether the series $\{\theta_t\}_{t=1}^n$ contains a change-point or not, so a testing step is called for prior to the estimation step.   Formally speaking,  given  a pre-specified threshold $K_n$, we declare the existence of a change-point when $$SN_n:=\max_{k=1,\cdots,n-1}T_n(k) > K_n,$$ 
and then estimate the single change-point via \eqref{sn_k}.
Otherwise, if $SN_n$ is below the threshold $K_n$, we declare no change-points.

\subsection{Multiple Change-Point Estimation}\label{subsec:SNCPmultiple}	

Section \ref{subsec:SNCPsingle} introduces how SNCP works in the single change-point setting. In this section, we further discuss its implementation for multiple change-point estimation. Compared with the single change-point setting, the main difficulty of multiple change-point estimation lies in how to isolate one change-point from another. In SNCP, this is achieved by a nested local-window approach. 

We first introduce some notations. Assume there are $m_0 \geq 0$ unknown number of change-points with $k_0=0<k_1<\cdots<k_{m_0}<n=k_{m_0+1}$ that partition $Y_t$ into $m_0+1$ stationary segments with constant quantity of interest $\theta^{(i)}$ in the $i$th segment, for $i=1,\cdots,m_0+1$.  In other words,
\begin{equation*}
	\theta_t = \theta^{(i)},~~ k_{i-1}+1 \leq t \leq k_i, ~\text{for}~ i = 1,\cdots,m_0+1.
\end{equation*}
Similar to the single change-point estimation framework, for $1 \leq t_1 < k < t_2 \leq n$, we define an SN based test statistic
\begin{equation}\label{eq:Tnt1kt2}
	T_n(t_1,k,t_2)=D_n(t_1,k,t_2)'V_n^{-1}(t_1,k,t_2)D_n(t_1,k,t_2), 
\end{equation}
where $D_n(t_1,k,t_2) = \frac{(k-t_1+1)(t_2-k)}{(t_2-t_1+1)^{3/2}}(\hat{\theta}_{t_1,k}-\hat{\theta}_{k+1,t_2})$, $V_n(t_1,k,t_2) = L_n(t_1,k,t_2) + R_n(t_1,k,t_2)$ and
\begin{eqnarray*}
	L_n(t_1,k,t_2) = \sum_{i=t_1}^{k} \frac{(i-t_1+1)^2(k-i)^2}{(t_2-t_1+1)^2(k-t_1+1)^2}(\hat{\theta}_{t_1,i}-\hat{\theta}_{i+1,k})^{\otimes 2},   \\
	R_n(t_1,k,t_2) = \sum_{i=k+1}^{t_2} \frac{(t_2-i+1)^2(i-1-k)^2}{(t_2-t_1+1)^2(t_2-k)^2}(\hat{\theta}_{i,t_2}-\hat{\theta}_{k+1,i-1})^{\otimes 2}.
\end{eqnarray*}
Here $T_n(t_1,k,t_2)$ plays the same role as $T_n(k)$ in \eqref{sn_k}, except for the fact that it is defined on the subsample $\{Y_t\}_{t=t_1}^{t_2}$. In other words, $T_n(t_1,k,t_2)=T_n(k)$ if $t_1=1$ and $t_2=n$. 

\begin{figure}[H]
	\centering
	\includegraphics[width=0.8\textwidth]{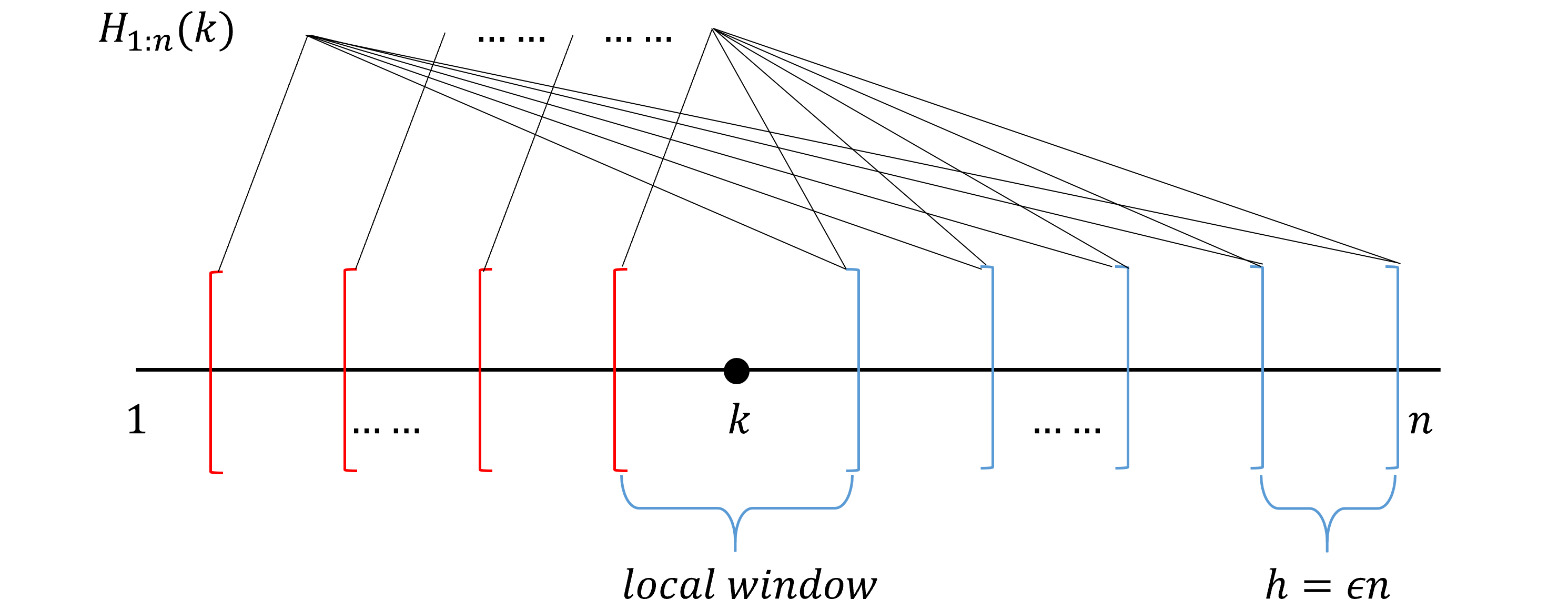}
	\caption{Graphical illustration of the nested local-windows in $H_{1:n}(k)$. Each pair of the red left bracket ${\color{red}[}$ and the blue right bracket ${]}$ represents a local-window in $H_{1:n}(k)$. }
	\label{fig:local_window}
\end{figure}

We now combine the SN framework with a nested local-window segmentation algorithm in \cite{zhao2021segmenting} for multiple change-point estimation.  For each $k$, instead of using the global statistic  $T_n(1,k,n)$ which is computed with all observations, we compute a maximal SNCP test statistic based on a collection of nested windows covering $k$. Specifically, we fix a small trimming parameter $\epsilon \in (0,1/2)$ and define the window size $h=\lfloor{n\epsilon}\rfloor$. For each $k=h,\cdots,n-h$, we define the nested local-window set $H_{1:n}(k)$ as
\begin{equation}\label{window_set}
	H_{1:n}(k) = \{ (t_1,t_2)|t_1=k-j_1h+1,j_1=1,\cdots,\lfloor{k/h}\rfloor;t_2=k+j_2h,j_2=1,\cdots,\lfloor{(n-k)/h}\rfloor. \}
\end{equation}
Note that for $k<h$ and $k>n-h$, we have $H_{1:n}(k)=\emptyset$. In Figure \ref{fig:local_window}, we plot a graphical illustration of the nested local-windows in $H_{1:n}(k)$, where local windows are constructed by combining  every pair of the red left bracket ${\color{red}[}$ and the blue right bracket ${]}$. 

For each $k=1,\cdots,n-1$, based on its nested local-window set $H_{1:n}(k)$, we define a maximal SN test statistic such that 
\begin{equation}\label{max_T}
	T_{1,n}(k) = \max_{(t_1,t_2) \in H_{1:n}(k)} T_n(t_1,k,t_2),
\end{equation}
where we set $\max_{(t_1,t_2) \in \emptyset} T_n(t_1,k,t_2) = 0$. 

Intuitively, with a sufficiently small trimming $\epsilon$, the nested local-window framework ensures that for a true change-point location, say $k^*$,  there exists some local window set denoted by $(t_1^*,t_2^*)$ containing $k^*$ as the only change-point. In other words, $k^*$ is isolated by the interval $(t_1^*,t_2^*)$ so that the procedure in the single change-point scenario as Section \ref{subsec:SNCPsingle} can be applied. This suggests that for at least one pair of $(t_1,t_2)$ in $H_{1:n}(k)$, $T_n(t_1,k^*,t_2)$ is large. The detection power is further enhanced by taking the maximum of these test statistics.

Based on the maximal test statistic $T_{1,n}(k)$ and a pre-specified threshold $K_n$,  SNCP  proceeds as follows. Starting with the full sample $\{Y_t\}_{t=1}^{n}$, we calculate $T_{1,n}(k), k=1,\cdots, n.$ Given that $\max_{k=1,\ldots,n} T_n(k)\leq K_n$, SNCP declares no change-point. Otherwise, SNCP sets $\widehat{k}=\arg\max_{k=1,\ldots,n} T_{1,n}(k)$ and we recursively perform SNCP on the subsample $\{Y_t\}_{t=1}^{\widehat{k}}$ and $\{Y_{t}\}_{t=\widehat{k}+1}^{n}$ until no change-point is declared.  Denote $W_{s,e}=\bigl\{(t_1,t_2)\big\vert s\leq t_1<t_2\leq e \bigl\}$ and $H_{s:e}(k)=H_{1:n}(k)\bigcap W_{s,e}$, which is the nested window set of $k$ on the subsample $\{Y_{t}\}_{t=s}^e$. Define the subsample maximal SN test statistic as $T_{s,e}(k)=\max\limits_{(t_1,t_2)\in H_{s:e}(k)}T_n(t_1,k,t_2).$ Algorithm \ref{alg:one} states the formal description of SNCP in multiple change-points estimation.

{ We note that SNCP shares some similarity with binary segmentation (BS) in the sense that both algorithms search for change-points in a \textit{sequential} fashion. However, they are also quite different. In particular, the SN test statistic in SNCP is computed over a set of nested local-windows instead of over a single interval. In contrast, in the classical change-point literature, BS is usually coupled with a global CUSUM statistic computed over the entire dataset $[1,n]$. As is documented in the literature \citep{shao2010self}, the main drawback of BS is its power loss under non-monotonic change, which is  caused by the use of the global CUSUM statistic coupled with the sequential search. However, due to the use of the nested local-window based SN test statistic, SNCP does not suffer from this power loss phenomenon as long as $\epsilon$ is chosen to be smaller than the minimum spacing between two change-points.  On the other hand, due to the  sequential  search nature, both SNCP and BS may encounter the multiple testing problem. In addition, their power may be lesser compared to a global search algorithm, such as dynamic programming or PELT, which is again due to the sequential nature of the search algorithm.
}
\begin{algorithm}
	\caption{SNCP procedures for multiple change-point  estimation}\label{alg:one}
	\textbf{Input:} Time series $\{Y_t\}_{t=1}^n$, threshold $K_n$, window size $h=\lfloor n\epsilon\rfloor$
	
	\textbf{Output:} Estimated change-points $\widehat{\mathbf{k}}=(\hat{k}_1,\cdots,\hat{k}_{\hat{m}})$
	
	\textbf{Initialization:} SNCP($1,n,K_n,h$), $\widehat{\mathbf k}=\emptyset$
	
	\textbf{Procedure:} SNCP($s,e,K_n,h$)
	
	\eIf{$e-s+1<2h$}
	{
		stop
	}{
		$\hat{k}^*=\arg\max_{k=s,\cdots,e}T_{s,e}(k)$;
		
		\eIf{$T_{s,e}(\hat{k}^\ast)\leq K_n$}
		{
			stop
		}{
			$\widehat{\mathbf{k}}=\widehat{\mathbf{k}}\cup\hat{k}^\ast$;
			
			run SNCP($s,\hat{k}^\ast,K_n,h$) and SNCP($\hat{k}^\ast+1,e,K_n,h$);
		}
	}
	
\end{algorithm}

\subsection{Multiple Change-Point Estimation for High-Dimensional Mean}\label{subsec:SNHD}
In this section, we modify the SNCP framework in Section \ref{subsec:SNCPmultiple} to design a new algorithm called SNHD for multiple change-point estimation in the mean vector of a high-dimensional time series.

Different from the subsample test statistic $T_n(t_1,k,t_2)$ used in SNCP for a low-dimensional time series, SNHD is designed based on the high-dimensional U-statistic proposed by \cite{wang2022inference}. Given a $p$-dimensional time series $\{Y_t\}_{t=1}^n$, we define the subsample contrast statistic as 
\begin{equation}\label{eq:DnU}
	D_{n}^{U}(t_1,k,t_2)=\sum_{\substack{t_1\leq j_1,j_3\leq k \\ j_1\neq j_3}} \sum_{\substack{k+1\leq j_2,j_4\leq t_2 \\ j_2\neq j_4}} (Y_{j_1}-Y_{j_2})^{\top}(Y_{j_3}-Y_{j_4}).
\end{equation}
Note that $D_{n}^{U}(t_1,k,t_2)$ is a two-sample U-statistic estimating the squared $L_2$-norm of the difference between the means  of $\{Y_t\}_{t=t_1}^k$ and $\{Y_t\}_{t=k+1}^{t_2}$ (up to some normalizing constant), { and therefore targets dense changes in high-dimensional mean.} The statistic in \eqref{eq:DnU} is only applicable to high-dimensional time series with temporal independence, and in the presence of temporal dependence, a trimming parameter needs to be introduced to alleviate the bias due to serial dependence; see \cite{wang2022inference}.  
Define the self-normalizer as 
\begin{equation}
	V_{n}^{U}(t_1,k,t_2)=\frac{1}{n}\Big[ \sum_{t=t_1+1}^{k-2} D_{n}^{U}(t_1,t,k)^2+\sum_{t=k+2}^{t_2-2}D_{n}^{U}(k+1,t,t_2)^2 \Big].
\end{equation}
The subsample SNHD test statistic at time point $k$, in the same spirit as \eqref{max_T}, is defined  as 
\begin{equation}
	T_{1,n}^{U}(k)=\max_{(t_1,t_2)\in H_{1:n}(k)}T_{n}^{U}(t_1,k,t_2),\quad  T_{n}^{U}(t_1,k,t_2)=D_{n}^{U}(t_1,k,t_2)^2/V_{n}^{U}(t_1,k,t_2),
\end{equation}
where $H_{1:n}(k)$ is the nested local-window set defined in \eqref{window_set}. 
With a pre-specified threshold $K_n^{U}$, a change-point is detected at $\hat{k}=\arg\max_{k=1,\cdots,n}T_{1,n}^U(k)$ if  $\max_{k=1,\cdots,n}T_{1,n}^{U}(k)$ is above $K_n^U$. For multiple change-point  estimation, SNHD proceeds similarly as SNCP  in Algorithm \ref{alg:one}. 

\subsection{Choice of Trimming Parameter $\epsilon$ and  Threshold $K_n$}\label{subsec:SNtheory}
For practical implementation of SNCP and SNHD, there are still two tuning parameters that one has to choose, namely the trimming parameter $\epsilon$ and the change-point detection threshold $K_n$. The choice of  $\epsilon$ reflects one's   belief of the minimum (relative) spacing between two consecutive change-points. This is usually set to be a small constant such as 0.05, 0.10, 0.15.  { The theoretical validity of our approach requires the minimum spacing between change-points to be of order $O(n)$, and opting for an overly small value of $\epsilon$ may result in sub-optimal performance in finite sample as the nested local-window may not contain sufficient observations. On the other hand, an overly large value of $\epsilon$ may increase the potential risk of under-estimating change-points if $\epsilon$ is larger than the minimum spacing between two true change-points. In practice, we recommend using 0.05 as a default value when no prior knowledge of minimum spacing between change-points is available. }

{  A nice feature of using SN is that the limiting distributions for SNCP or SNHD under the no change-point scenario are pivotal and furthermore reflect the impact of the choice of $\epsilon$, see Theorem 3.1 in \cite{zhao2021segmenting}, and Section S.2.9 in \cite{zhao2021v1}, respectively.} Since $K_n$ and $K_n^U$ are used to balance one's tolerance of type-I and type-II errors, this implies that we can  choose $K_n$ and $K_n^U$ as the  $q\times 100\%$ quantiles (i.e.\ the critical value) of the limiting null distribution with $q$ typically set as 0.9, 0.95, 0.99. { The threshold value $K_n$ also increases with the dimension of the quantity $\theta$, and we refer to Table 1 in \cite{zhao2021segmenting} for details.} In the \pkg{SNSeg} package, we offer users a wide range of $\epsilon$ and $q$ to choose from. Details are given in the following section.

{ In Section \ref{subsec:execution-time}, we further conduct a sensitivity analysis, which suggests that the performance of SNCP in general is robust w.r.t.\ the choice of $(\epsilon,K_n)$.}

\section{The SNSeg Package}\label{sec:SNSeg}
In this section, we introduce the functions within the \pkg{SNSeg} package for multiple change-point estimation. In particular,  \texttt{SNSeg\_Uni()} in  Section \ref{subsec:SNSegUni} implements the SNCP procedure of change-point estimation for a univariate time series with changes in a single or multiple parameters, such as mean, variance, auto-correlations, quantiles or even their combinations. { It can also be implemented for detecting change-points in other quantities, with a user-defined function as input.} The function \texttt{SNSeg\_Multi()} in  Section  \ref{subsec:SNSegMulti}   utilizes the SNCP algorithm for change-point estimation in mean or covariance matrix of a multivariate time series. In  Section \ref{subsec:SNSegHD},  the function \texttt{SNSeg\_HD()} estimates change-points in mean of a high-dimensional time series using the SNHD procedure. In addition to these major functions for change-point estimation, we further introduce \texttt{max\_SNsweep()} in Section  \ref{subsec:teststatistic}, which helps obtain the SN test statistics and create a segmentation plot based on the output of the above functions. {Followed by the graphical options, the function \texttt{SNSeg\_estimate()} generates parameter estimates within each segment separated by the estimated change-points.}

\subsection{SNCP for Univariate Time Series}\label{subsec:SNSegUni}

For a univariate time series, change-point estimation in a single or multiple functionals can be implemented through the function \texttt{SNSeg\_Uni()}. This function is also capable of detecting change-points associated with the change in correlation between bivariate time series.  The \textbf{R} code is given as:
\begin{example}
	SNSeg_Uni(ts, paras_to_test, confidence = 0.9, grid_size_scale = 0.05,
	grid_size = NULL, plot_SN = TRUE, est_cp_loc = TRUE)
\end{example}
It takes the following input arguments.
\begin{itemize}
	\item \texttt{ts}: Input time series $\{Y_t\}_{t=1}^n$, i.e., a univariate time series expressed as a numeric vector with length $n$. However, when the argument \texttt{paras\_to\_test} is specified as \texttt{bivcor}, which stands for the correlation between bivariate time series, the input \texttt{ts} must be an $n\times 2$ matrix.
	
	\item \texttt{paras\_to\_test}: The parameters that SNCP aims to examine, which are presented as a string, a number, a combination of both, {or a user-defined function that defines a specific functional}. Available options of \texttt{paras\_to\_test} include:
	\begin{itemize}
		\item \texttt{"mean"}: The function performs change-point estimation on the mean of the time series.
		\item \texttt{"variance"}: The function performs change-point estimation on the variance.
		\item \texttt{"acf"}: The function performs change-point estimation on the autocorrelation of order 1.
		\item \texttt{"bivcor"}: The function performs change-point estimation on the bivariate correlation.
		\item A numeric quantile value within the range (0,1): The function performs change-point estimation on the quantile level specified by the numeric value.
		\item A vector containing characters \texttt{"mean"}, \texttt{"variance"}, \texttt{"acf"}, and one or more numerical quantile levels. Therefore, \texttt{SNSeg\_Uni()} is capable of estimating change-points in either a single parameter or a combination of multiple parameters.
		\item {A user-defined \texttt{R} function that returns a numeric value. Existing functions in \textbf{R} such as $\texttt{mean}()$ and $\texttt{var}()$ can also be used. This option provides additional flexibility for the users to define a specific functional that they are interested in and is not covered by our built-in options. The input argument \texttt{paras\_to\_test} should possess the form of \texttt{function(...)\{\}}.}
	\end{itemize}	
	\item \texttt{confidence}: A numeric value that specifies the confidence level of the SN test. Available choices of confidence levels contain \texttt{0.9}, \texttt{0.95}, \texttt{0.99}, \texttt{0.995} and \texttt{0.999}. It automatically obtains the threshold ($K_n$, the critical value) corresponding  to the input confidence level. 
	% Conversely, if the change-points are estimated based on multiple parameters, the critical values can be derived from the table \texttt{critical\_values\_multi}.
	The default value of \texttt{confidence} is set at 0.9.
	
	\item \texttt{grid\_size\_scale}: A numeric value that specifies the trimming parameter $\epsilon$ and only in use if \texttt{grid\_size = NULL}. Available choices include  0.05, 0.06, 0.07, 0.08, 0.09, 0.1, 0.11, 0.12, 0.13, 0.14, 0.15, 0.2, 0.25, 0.3, 0.35, 0.4, 0.45 and 0.5.  The default value of \texttt{grid\_size\_scale} is 0.05.  
	\begin{itemize}
		\item In the function, any input less than 0.05 will be set to exactly 0.05 and similarly, any input greater than 0.5 will be set to 0.5. In such case, a warning that "Detected the grid\_size\_scale is greater than 0.5" or "less than 0.05" will be generated. 
	\end{itemize}
	
	\item \texttt{grid\_size}: A numeric value that specifies the local window size $h$. It should be noted that $h=\lfloor n\times\epsilon\rfloor$, i.e., \texttt{grid\_size}$= \lfloor n\times \texttt{grid\_size\_scale}\rfloor$. By default, the value of \texttt{grid\_size} is set to NULL, and the function computes the critical value $K_n$ using the argument \texttt{grid\_size\_scale}.  However, users have the option to set the \texttt{grid\_size} manually, in which case the function computes the corresponding \texttt{grid\_size\_scale} via dividing \texttt{grid\_size} by $n$, and then determines the critical value using this computed \texttt{grid\_size\_scale} value.
	
	\item \texttt{plot\_SN}: A Boolean value that specifies whether to plot the time series or not. The default setting is TRUE. 
	
	\item \texttt{est\_cp\_loc}: A Boolean value that specifies whether to plot a red vertical line for each estimated change-point. The default setting is TRUE.
\end{itemize}

The function \texttt{SNSeg\_Uni()} provides users with flexibility by allowing them to select parameter types that they want to target at. Additionally, users can specify the window size or choose an appropriate value for $\epsilon$ to achieve the desired theoretical critical value. 
{However, if $\epsilon$ happens to be larger than the true minimum spacing between change-points, the function carries the risk of missing some of them. This limitation  also applies to the functions \texttt{SNSeg\_Multi()} and \texttt{SNSeg\_HD()}. In practice, we suggest $\epsilon=0.05$ with no prior knowledge.
	% A comprehensive discussion on this limitation and the rationale behind selecting $\epsilon$ is provided in Section \ref{sec:additional}.
} If the calculated trimming parameter $\epsilon$ by any of the two arguments falls within the range [0.05, 0.5], but is not in the pre-specified set of available values for \texttt{grid\_size\_scale}, the function performs a linear interpolation by identifying two nearest \texttt{grid\_size\_scale} that are below and above the calculated $\epsilon$ and then computes the weighted average of the corresponding critical values $K_n$. The resulting interpolated value is used as the final critical value for the SN test.

When called, \texttt{SNSeg\_Uni()} returns {an S3 object of class \texttt{SNSeg\_Uni}} containing the following entries.
\begin{itemize}
	\item \texttt{ts}: The input time series \texttt{ts}.
	\item \texttt{paras\_to\_test}: The parameter(s) examined in change-point estimation. 
	\item \texttt{grid\_size}: A numeric value of the local window size $h$.
	\item \texttt{SN\_sweep\_result}: A list of $n$ matrices where the $k$th matrix stores the SN test statistic $T_n(t_1,k,t_2)$ computed for all $(t_1,t_2)\in H_{1:n}(k)$ as in \eqref{max_T}. In particular, the $k$th matrix consists of four columns: 1.\ the SN test statistic $T_n(t_1,k,t_2)$ computed via \eqref{eq:Tnt1kt2}; 2.\ the location $k$; 3.\ the left endpoint $t_1$; and 4.\ the right endpoint $t_2$. 
	\item \texttt{est\_cp}: A numeric vector containing the locations of the estimated change-points.
	\item \texttt{confidence}: The confidence level of the SN test.
	\item \texttt{critical\_value}: The critical value of the SN test given $\epsilon$ and the confidence level.
\end{itemize}

It is worth noting that the output of the function \texttt{SNSeg\_Uni()} can  serve as an input of the function \texttt{max\_SNsweep()} (to be described in Section \ref{subsec:teststatistic}) to generate a segmentation plot for SN test statistics, and the same also holds for functions \texttt{SNSeg\_Multi()} and \texttt{SNSeg\_HD()}. 
{Additionally, S3 objects of class \texttt{SNSeg\_Uni} are supported by  \texttt{print()},  \texttt{summary()} and \texttt{plot()} functions. The S3 function \texttt{print()} can be used to display the estimated change-points,   \texttt{summary()} presents information such as change-point locations and other details listed in the output of \texttt{SNSeg\_Uni()}, and \texttt{plot()} facilitates the generation of time series segmentation plots, providing an alternative option to the argument \texttt{plot\_SN = TRUE} for users. These functions can also be applied to outputs of functions \texttt{SNSeg\_Multi()} and \texttt{SNSeg\_HD()}, which will be introduced below.}

To illustrate, in the following, we present examples demonstrating multiple change-point estimation in both single and multiple parameters. 

\noindent{\bf Example 1:  variance change in univariate time series}

We start with the example of the change-point model (V1) in Section S.2.5 in the supplement of \cite{zhao2021segmenting}, where two variance changes occur at  $k=400$ and 750, respectively. Specifically, $$
(\mathrm{V} 1): \quad Y_t= \begin{cases}0.5 Y_{t-1}+\epsilon_t, & t \in[1,400], \\ 0.5 Y_{t-1}+2 \epsilon_t, & t \in[401,750], \\ 0.5 Y_{t-1}+\epsilon_t, & t \in[751,1024],\end{cases}
$$
where $\epsilon_t$ is a sequence of i.i.d. $N(0,1)$ random variables.

We set \texttt{grid\_size\_scale} at $\epsilon=0.05$, which corresponds to a \texttt{grid\_size} of $\lfloor 0.05*1024\rfloor=51$, and set \texttt{confidence} at $90\%$. Subsequently, we visualize the input time series by setting \texttt{plot\_SN} as \texttt{TRUE}, and generate an SN test statistics segmentation plot using the \texttt{max\_SNsweep()} function (to be introduced in Section \ref{subsec:teststatistic}). 

\begin{example}
	# Generate model (V1)
	set.seed(7)
	ts <- MAR_Variance(reptime = 1, type = "V1") # generate model (V1)
	par(mfcol = c(2, 1), mar = c(4, 2.5, 2.5, 0.5))
	
	# SNCP in the change of variance
	result1 <- SNSeg_Uni(ts, paras_to_test = "variance", confidence = 0.9,
	grid_size_scale = 0.05, grid_size = NULL, plot_SN = TRUE, 
	est_cp_loc = TRUE)
	
	# Segmentation plot for SN-based test statistics
	SNstat1 <- max_SNsweep(result1, plot_SN = TRUE, est_cp_loc = TRUE, critical_loc = TRUE)
\end{example}
The estimated locations of the change-points $\widehat{\mathbf k}$, the local window size $h$, and the critical value $K_n$ used can be accessed using the following commands:

\begin{example}
	result1$est_cp
	[1] 411 748
	result1$grid_size
	[1] 51
	result1$critical_value
	[1] 141.8941
\end{example}

The estimated change-point locations are $\hat{k}=411$ and 748 with a local window size of $h=51$ and critical value of $K_n=141.8941$ when setting the trimming parameter to $\epsilon=0.05$ and the confidence level to $90\%$. It is clear that the estimated change-points align closely with the true change-points, which demonstrates the accuracy of the SNCP procedure.  {The above outputs can also be obtained via the S3 methods \texttt{summary()} and \texttt{print()}, and are shown in the following commands: 
	% \textcolor{red}{Which example is this? Why the cp is 1018 and 1487?} \textcolor{purple}{Shubo: This is a typo. I have re-run the code and updated the correct output.}

	\begin{example}
		# S3 method: print
		print(result1)
		#> The detected change-point location(s) are 411,748
		# S3 method: summary
		summary(result1)
		#> There are 2 change-points detected at 90th confidence level based on the change in 
		#> the single variance parameter.
		#> 
		#> The critical value of SN-based test is 141.8941189
		#> 
		#> The detected change-point location(s) are 411,748 with a grid_size of 51
	\end{example} 
}

\begin{figure}[!h]
	\centering
	\includegraphics[scale=0.5]{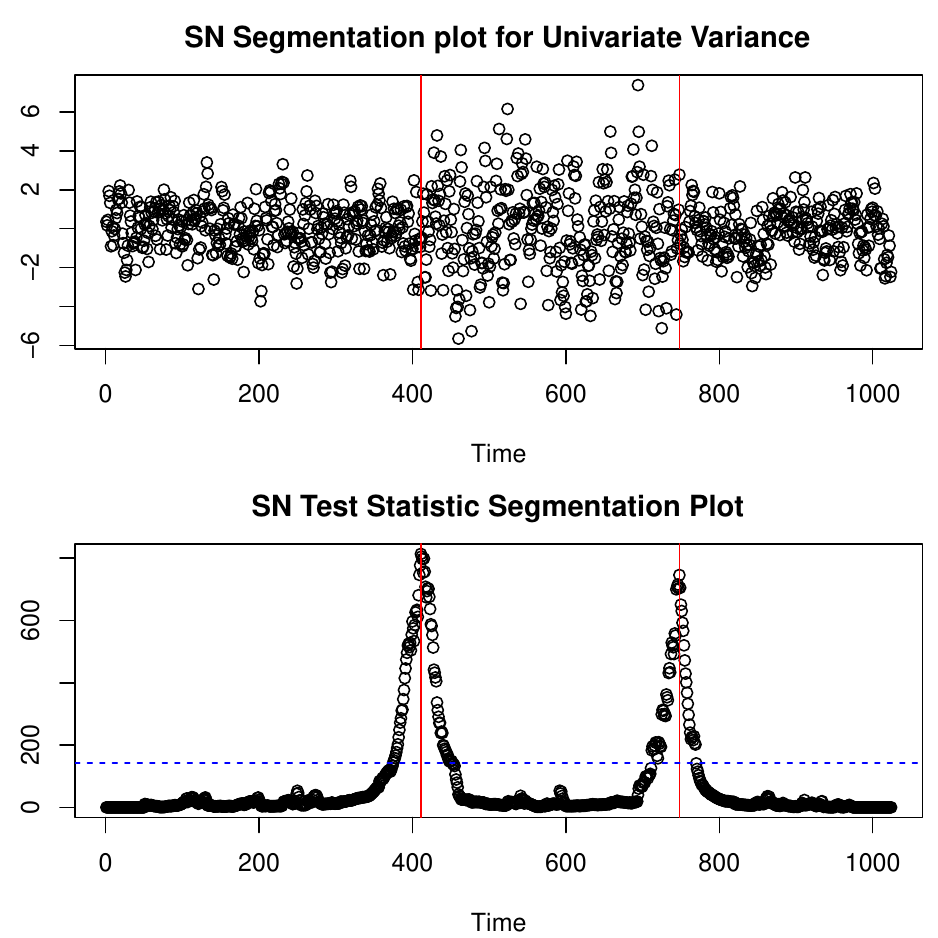}
	\caption{An example of simulated time series of model (V1). The upper panel illustrates the detection of change-points for the input time series, and the lower panel displays the segmentation of the SN test statistics using the estimated change-points. The detected change-point locations are indicated by a red vertical line and the critical value is represented by a blue horizontal line.}
	\label{fig:SN-uni-single-var}
\end{figure}

Figure \ref{fig:SN-uni-single-var} displays segmentation plots for the input time series and the SN test statistics regarding the changes in univariate variance. It reveals that the SN statistics associated with the detected change-points surpass the critical value, and can be deemed as plausible changes. {The upper plot (SN segmentation plot of the time series) can also be achieved through the command \texttt{plot(result1)}.}

{In addition to the summary statistics and SN based segmentation plots, the function \texttt{SNSeg\_estimate()} provides parameter estimates within each segment that is separated by the estimated change-points. To illustrate this, we apply the following command to the same example.

	\begin{example}
		SNSeg_estimate(SN_result = result1)
		
		# output
		$variance
		[1] 1.438164 5.065029 1.341965
		
		attr(,"class")
		[1] "SNSeg_estimate"
	\end{example}
	
}

We can also manually specify the value of \texttt{grid\_size} to calculate the critical value $K_n$ and estimate change-points. The function \texttt{SNSeg\_Uni()} is applied to the same time series, with the only difference being that the window size $h$ is set to 102, which corresponds to \texttt{grid\_size\_scale=0.1}, instead of \texttt{NULL}. We further set \texttt{confidence} at 90\%. The estimated change-points and the critical value $K_n$ can be obtained using the following commands:

\begin{example}
	# SNCP in the change of variance with a different grid_size and confidence level
	result2 <- SNSeg_Uni(ts, paras_to_test = "variance", confidence = 0.9, 
	grid_size_scale = 0.05, grid_size = 102, plot_SN = FALSE,
	est_cp_loc = FALSE)
	result2$est_cp
	[1] 411 744
	result2$grid_size
	[1] 102
	result2$critical_value
	[1] 111.1472
\end{example}

Note that since \texttt{grid\_size} is not \texttt{NULL}, the argument \texttt{grid\_size\_scale = 0.05} will be ignored by the function \texttt{SNSeg\_Uni()}. Interestingly, though we use quite different window size $h=102$, the estimated change-points are almost the same as before, which suggests the robustness of SNCP. The critical value differs from the previous example, due to the variation in the window size $h$ (or equivalently $\epsilon$). In other words,the threshold $K_n$ used in SNCP reflects the influence of the chosen window size $h$ (or $\epsilon$), which makes the change-point detection more robust and accurate.

{
	\noindent{\bf Example 2:  second moment change in univariate time series with a user-defined function}

	In addition to the built-in parameter choices, the function \texttt{SNSeg\_Uni()} allows users to customize the input parameter using their own function. For instance, if users are interested in examining changes in the second moment, they can create a function that yields the mean square as a numeric value and designate this function to the input argument for \texttt{paras\_to\_test}. To illustrate,  we consider the model (V1) from \textbf{Example 1}. The function \texttt{SNSeg\_Uni()} is utilized with the default input configuration, except that we now assess the change in the second moment of the data. The specified \texttt{paras\_to\_test}, along with the execution time and the resultant estimated change-points, can be acquired using the following commands:}

\begin{example}
	# define a function for paras_to_test
	# change in 2nd moment
	second_moment <- function(ts){
		result <- mean(ts^2)
		return(result)
	}
	start.time <- Sys.time()
	result.general <- SNSeg_Uni(ts, paras_to_test = second_moment, confidence = 0.9,
	grid_size_scale = 0.05, grid_size = NULL,
	plot_SN = FALSE, est_cp_loc = TRUE)
	end.time <- Sys.time()
	as.numeric(difftime(end.time,start.time)) # execution time (in minutes)
	result.general$est_cp # change-point estimates    
	
	# Output 
	> as.numeric(difftime(end.time,start.time)) # execution time (in minutes)
	[1] 22.0148
	> result.general$est_cp # change-point estimates
	[1] 412 749
\end{example}

{As evident from the above results, SNCP detected two change-points at $\hat{k}=412,749$ when examining changes in the second moment. The estimated change-points are close to the locations of the true change-point locations, 400 and 750, respectively.}

{To illustrate the computational efficiency of the built-in choices of \texttt{paras\_to\_test} (i.e., \texttt{"mean"}, \texttt{"variance"}, etc.) another example is given to detect changes in variance but using the user-defined functional \texttt{{var()}}, which is then compared with the built-in option \texttt{paras\_to\_test = "variance"} in terms of the execution time. The comparison result can be accessed using the following commands:}

\begin{example}
	start.time <- Sys.time()
	result1 <- SNSeg_Uni(ts, paras_to_test = "variance", confidence = 0.9,
	grid_size_scale = 0.05, grid_size = NULL, plot_SN = TRUE,
	est_cp_loc = TRUE)
	end.time <- Sys.time()
	difftime(end.time,start.time)) # built-in parameter time
	
	# user defined variance
	paras_to_test <- function(ts){
		var(ts)
	}
	start.time <- Sys.time()
	result.general <- SNSeg_Uni(ts, paras_to_test = paras_to_test, confidence = 0.9,
	grid_size_scale = 0.05, grid_size = NULL,
	plot_SN = FALSE, est_cp_loc = TRUE)
	end.time <- Sys.time()
	difftime(end.time,start.time) # general functional parameter time
	c(result1$est_cp,result.general$est_cp) 
	
	# output
	> difftime(end.time,start.time) # built-in parameter time
	Time difference of 4.702668 secs
	> difftime(end.time,start.time) # general functional parameter time
	Time difference of 13.39276 mins
	> result1$est_cp # built-in parameter estimate 
	[1] 411 748
	> result.general$est_cp # general functional estimate
	[1] 411 748
\end{example}

{Both methods can accurately estimate change-points, but utilizing the built-in parameter significantly reduces computation time compared to using user-defined function.  The former method optimizes efficiency by leveraging the linear structure of variance calculation and is implemented via dynamic programming with the \texttt{cumsum()} function. In contrast, the latter method, employing a user-defined function, does not utilize the linear structure of variance (since it takes a general functional as an input which may not have a specific structure) and instead recursively calculates all subsample variances, leading to redundant calculations and increased computational time. }

% We  note that the multiple-parameters setting is only applicable to univariate time series, and hence the option for bivariate correlation (\texttt{bivcor}) is not available. 

\noindent{\bf Example 3:  multiple-parameter change in univariate time series}

In addition to identifying changes in a single parameter, \texttt{SNSeg\_Uni()} also allows for estimating change-points by simultaneously combining information across multiple parameters. This can be done by modifying the \texttt{paras\_to\_test} argument. For example, users can specify \texttt{paras\_to\_test = c("mean", "acf", 0.6, 0.9)} to simultaneously detect changes in mean, autocorrelation, 60\% and 90\% quantile of the input time series.
%We consider the simulated univariate time series of model (MP2) in Section 4.4 of \cite{zhao2021segmenting}, where the true change-points are located at $k=333$ and 667. In particular, 
%$$
%(\mathrm{MP} 2): Y_t= \begin{cases}\epsilon_t, & t \in[1,333] \\1.6 \epsilon_t, & t \in[334,667] \\ \epsilon_t, & t \in[668,1000],\end{cases}
%$$
%where $\{\epsilon_t\}_{t=1}^n$ is a sequence of  i.i.d. $N(0,1)$ random variables.

We consider the simulated univariate time series of model (MP1) in Section 4.4 of \cite{zhao2021segmenting}, where the true change-points are located at $k=333$ and 667. In particular, 
$$
(\mathrm{MP} 1): Y_t= \begin{cases}X_t, & t \in[1,333] \\ F^{-1}\left(\Phi\left(X_t\right)\right), & t \in[334,667] \\ X_t, & t \in[668,1000],\end{cases}
$$
where $\{X_t\}_{t=1}^n$ follows an AR(1) process with $X_t=0.2 X_{t-1}+\sqrt{1-\rho^2}\epsilon_t$, $\epsilon_t$ is a sequence of  i.i.d.  $N(0,1)$ random variables, $\Phi(\cdot)$ denotes the CDF of $N(0,1)$, and $F(\cdot)$ denotes a mixture of a truncated normal and a generalized Pareto distribution~(GPD). In particular, $F(x)=0.5F_1(x)+0.5F_2(x)$, where $F_1(x)=2\Phi(x), x\leq 0$ is a standard normal distribution truncated at 0 and $F_2(x)=1-(1+\xi(x-\mu)/\sigma)_+^{-1/\xi}$ is a GPD distribution with the location parameter $\mu=0$, scale parameter $\sigma=2$ and tail index $\xi=0.125$. Note that $F^{-1}(q)=\Phi^{-1}(q)$ for $q\leq 0.5$ and $F^{-1}(q)\neq \Phi^{-1}(q)$ for $q>0.5.$ 

To showcase the versatility of SNCP, we first detect change-points based on the 90\% quantiles, where we set the \texttt{grid\_size\_scale} at $0.1$ and \texttt{confidence} at 0.9. 
\begin{example}
	set.seed(7)
	require(truncnorm)
	require(evd)
	mix_GauGPD <- function(u, p, trunc_r, gpd_scale, gpd_shape) {
		# function for generating a mixture of truncated normal + GPD
		indicator <- (u < p)
		rv <- rep(0, length(u))
		rv[indicator > 0] <- qtruncnorm(u[indicator > 0] / p, a = -Inf, b = trunc_r)
		rv[indicator <= 0] <- qgpd((u[indicator <= 0] - p) / (1 - p), loc = trunc_r, 
		scale = gpd_scale, shape = gpd_shape)
		return(rv)
	}
	
	# Generate model (MP1)
	n <- 1000
	cp_sets <- c(0, 333, 667, 1000)
	rho <- 0.2
	ts <- MAR(n, 1, rho) * sqrt(1 - rho ^ 2) # generate AR(1)
	trunc_r <- 0
	p <- pnorm(trunc_r)
	gpd_scale <- 2
	gpd_shape <- 0.125
	ts[(cp_sets[2] + 1):cp_sets[3]] <-
	mix_GauGPD(u = pnorm(ts[(cp_sets[2] + 1):cp_sets[3]]), p, trunc_r, gpd_scale, gpd_shape)
	
	# SNCP in the change of 90% quantile
	result_q9 <- SNSeg_Uni(ts, paras_to_test = c(0.9), confidence = 0.9, 
	grid_size_scale = 0.1, plot_SN = FALSE, est_cp_loc = FALSE)
	# Output
	result_q9$est_cp
	[1] 332 667
	result_q9$grid_size
	[1] 100
	result_q9$critical_value
	[1] 110.9993
\end{example}

As observed, the estimated change-points take place at $\hat{k}=332$ and 667, which are close to the locations of the true change-points. We can further use SNCP to examine if there is any change in the variance for the same time series.

\begin{example}
	# SNCP in the change of variance
	result_v <- SNSeg_Uni(ts, paras_to_test = c('variance'), confidence = 0.9, 
	grid_size_scale = 0.1, plot_SN = FALSE, est_cp_loc = FALSE)
	# Output
	result_v$est_cp
	[1] 329 665
	result_v$grid_size
	[1] 100
	result_v$critical_value
	[1] 110.9993
\end{example}

The estimated change-points in variance take place at $\hat{k}=329$ and 665, which are close to the estimated change-points in the 90\% quantile. To reconcile the two sets of estimated change-points, we can further examine changes in variance the 90\% quantile simultaneously using SNCP, which gives a final estimate of 331 and 667.

\begin{example}
	# SNCP in the change of variance and 90% quantile
	result_q9v <- SNSeg_Uni(ts, paras_to_test = c(0.9, 'variance'), confidence = 0.9, 
	grid_size_scale = 0.1, plot_SN = TRUE, est_cp_loc = TRUE)
	# Output
	result_q9v$est_cp
	[1] 331 667
	result_q9v$grid_size
	[1] 100
	result_q9v$critical_value
	[1] 167.4226
\end{example}

% { Figure \ref{fig:SN-uni-multiple-parameters} plots the time series (MP1) and estimated change-points.
	
	% \begin{figure}[H]
		% 	\centering
		% 	\includegraphics[scale=0.5]{multiple-parameters.eps}
		% 	\caption{An example of using the change in multiple parameters for model (MP1). The detected change-point locations are indicated by a red vertical line}
		% 	\label{fig:SN-uni-multiple-parameters}
		% \end{figure}}

{
	For practitioners, how to further identify which component(s) in “paras\_to\_test”
	changes is an interesting question. A natural strategy is as follows. For each detected change-point $\hat{\tau}$, we first construct a local window centered around it, e.g.\ $[\hat{\tau}-\epsilon n,\hat{\tau}+\epsilon n]$, where the local window should only contain a single change-point (with high probability). Within the local window, we then apply \texttt{SNSeg\_Uni}() for each parameter in “\texttt{paras\_to\_test}” and test if it changes.
}

\subsection{SNCP for Multivariate Time Series}\label{subsec:SNSegMulti}

The SN based change-point estimation for multivariate time series can be implemented via the function \texttt{SNSeg\_Multi()}. In particular, SNSeg\_Multi() allows change-point detection in multivariate means or covariance matrix of the input time series. The \textbf{R} code is given as:
\begin{example}
	SNSeg_Multi(ts, paras_to_test = "mean", confidence = 0.9, grid_size = NULL, 
	grid_size_scale = 0.05, plot_SN = FALSE, est_cp_loc = TRUE)
\end{example}

{The input arguments of   \texttt{confidence}, \texttt{grid\_size}, \texttt{grid\_size\_scale} and \texttt{est\_cp\_loc} are the same as those in the function \texttt{SNSeg\_Uni()}, and the difference lies in \texttt{ts}, \texttt{plot\_SN} and \texttt{paras\_to\_test}.}
\begin{itemize}
	\item \texttt{ts}: Input time series $\{Y_t=(Y_{t1},\cdots,Y_{tp})\}_{t=1}^n$ as a matrix, i.e., a multivariate time series represented as a matrix with $n$ rows and $p$ columns, where each column is a univariate time series. The dimension $p$ for \texttt{ts} should be at least 2.
	\item \texttt{paras\_to\_test}: A string that specifies the parameter that SNCP aims to examine. Available options of \texttt{paras\_to\_test} include:
	\begin{itemize}
		\item \texttt{"mean"}: The function performs change-point estimation on the mean of the multivariate time series.
		\item \texttt{"covariance"}: The function performs change-point estimation on the covariance matrix of the multivariate time series.   
	\end{itemize}
	\item {\texttt{plot\_SN}: A Boolean value that specifies whether to generate time series segmentation plot or not. \texttt{SNSeg\_Multi} returns a plot for each individual time series if \texttt{plot\_SN = TRUE}. }
\end{itemize}

When necessary, the function \texttt{SNSeg\_Multi()} applies the same linear interpolation rule as \texttt{SNSeg\_Uni()} to determine the critical value for the SN test. When called, \texttt{SNSeg\_Multi()} returns {an S3 object of class \texttt{SNSeg\_Multi}} comprising the \texttt{grid\_size}, \texttt{SN\_sweep\_result}, \texttt{est\_cp}, \texttt{confidence} and \texttt{critical\_value}, which have already been described in the context of the function \texttt{SNSeg\_Uni()}. It also generates plots for each time series when \texttt{plot\_SN = TRUE}.
{ Similar to \texttt{SNSeg\_Uni},  S3 objects of class \texttt{SNSeg\_Multi} are also supported by \texttt{print()},  \texttt{summary()} and \texttt{plot()} functions. }

% We note that \texttt{SNSeg\_Multi()} does not include the option to plot the input multivariate time series \texttt{ts}.

\noindent{\bf Example 4:  mean change in multivariate time series}

We consider model (M2) in Section 4.2 of \cite{zhao2021segmenting}, which is generated by 
$$
(\mathrm{M} 2):  Y_t= \begin{cases}-3 / \sqrt{5}+X_t, & t \in[1,75],[526,575], \\ 0+X_t, & t \in[76,375],[426,525],[576,1000], \\ 3 / \sqrt{5}+X_t, & t \in[376,425] .\end{cases}
$$
where $X_t$ is a 5-dimensional VAR(1) process with $X_t = 0.5X_{t-1} + \epsilon_t$, and $\epsilon_t$ is a sequence of i.i.d.  $N(0, \mathbf I_5)$ random vectors.

The five true change-points occur at $k=75,375,425,525$ and 575. We analyze it by examining the change in multivariate means using \texttt{grid\_size\_scale} at $0.05$ and \texttt{confidence} at 0.9.   The code implementation is as follows:
\begin{example}
	# Generate model (M2)
	set.seed(7)
	d <- 5
	n <- 1000
	cp_sets <- c(0, 75, 375, 425, 525, 575, 1000)
	mean_shift <- c(-3, 0, 3, 0, -3, 0) / sqrt(d)
	rho_sets <- 0.5
	sigma_cross <- list(diag(d))
	ts <- MAR_MTS_Covariance(n, 1, rho_sets, cp_sets = c(0, n), sigma_cross)[[1]] # generate VAR(1)
	no_seg <- length(cp_sets) - 1
	
	for (index in 1:no_seg) { # Mean shift
		tau1 <- cp_sets[index] + 1
		tau2 <- cp_sets[index + 1]
		ts[, tau1:tau2] <- ts[, tau1:tau2] + mean_shift[index]
	}
	
	par(mfrow=c(2,3))
	# SNCP in the change of multivariate mean
	result_multimean <- SNSeg_Multi(ts, paras_to_test = "mean", confidence = 0.9,
	grid_size_scale = 0.05, plot_SN = TRUE,
	est_cp_loc = TRUE)
	
	# Output
	result_multimean$est_cp
	[1] 80 373 423 526 576
	result_multimean$grid_size
	[1] 50
	result_multimean$critical_value
	[1] 415.8649
\end{example}

The estimated change-points are $\hat{k}=80,373,423,526$ and 576 with a window size $h=50$ and a critical value of 415.8649. This result again closely aligns with the true change-point locations. The output of \texttt{SNSeg\_Multi()} also allows for the use of function \texttt{max\_SNsweep()} for plotting the segmentation of the SN test statistics. The code implementation is as follows:
\begin{example}
	SNstat_multimean <- max_SNsweep(result_multimean, plot_SN = TRUE, est_cp_loc = TRUE,
	critical_loc = TRUE)
	plot(ts[1, ], main = 'SN Segmentation Plot for the First Time Series')
	abline(v = result_multimean$est_cp, col = 'red')
\end{example}

\begin{figure}[h]
	\centering
	\includegraphics[scale=0.5]{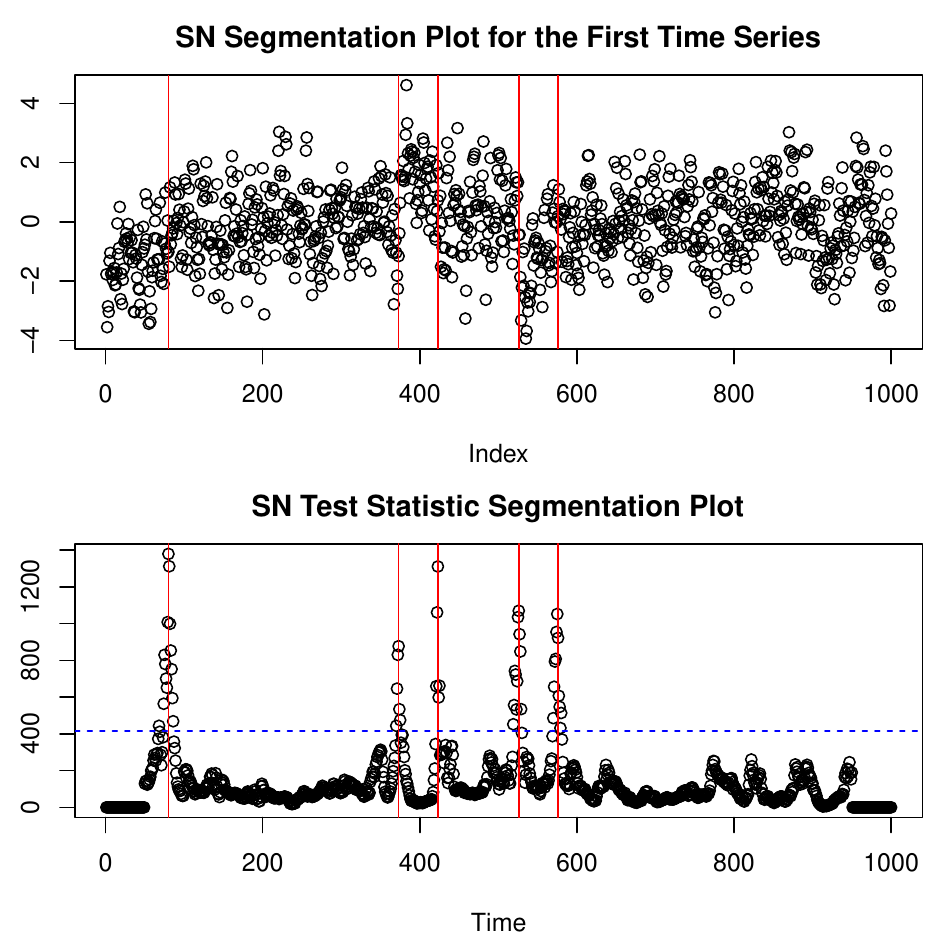}
	\caption{The segmentation of the SN test statistics using the estimated change-points. The detected change-point locations are indicated by a red vertical line and the critical value is represented by a blue horizontal line.}
	\label{fig:SN-multi-mean}
\end{figure}

Figure \ref{fig:SN-multi-mean} plots the associated SN test statistics and estimated change-points. For illustration, we also plot the first time series $\{Y_{t,1}\}_{t=1}^n$ along with the estimated change-points.

\subsection{SNHD for High-Dimensional Time Series}\label{subsec:SNSegHD}
The function \texttt{SNSeg\_HD()} is specifically designed to estimate change-points in the mean functional of  high-dimensional time series. The \textbf{R} code is given as:

\begin{example}
	SNSeg_HD(ts, confidence = 0.9, grid_size_scale = 0.05, grid_size = NULL,
	plot_SN = FALSE, est_cp_loc = TRUE, ts_index = c(1:5)
\end{example}

Its input arguments are the same as the function \texttt{SNSeg\_Multi()} except for the followings:

\begin{itemize}
	\item \texttt{ts}: The dimension of the input time series \texttt{ts} should be at least 10 to ensure a decent finite sample performance of the asymptotic theory.
	\item \texttt{plot\_SN}: {A Boolean value that specifies whether to return a plot for individual time series.}    
	\item \texttt{ts\_index}: {A positive integer or a vector of positive integers that specifies which individual time series to plot given \texttt{plot\_SN = TRUE}. The default value is \texttt{c(1:5)}, and under the default setting, the function will plot the first 5 time series.}
\end{itemize}

When called, \texttt{SNSeg\_HD()} returns {an S3 object of class \texttt{SNSeg\_HD}} containing \texttt{grid\_size}, \texttt{SN\_sweep\_result}, \texttt{est\_cp}, \texttt{confidence} and \texttt{critical\_value} that are similar to those described by \texttt{SNSeg\_Uni()} and \texttt{SNSeg\_Multi()}. {It also generates a plot for the time series specified by the argument \texttt{ts\_index} when \texttt{plot\_SN = TRUE}. Additionally, S3 objects of class \texttt{SNSeg\_HD} are supported by \texttt{plot()}, \texttt{summary()} and \texttt{print()} functions. Similar to the core function \texttt{SNSeg\_HD()}, the \texttt{plot()} method incorporates the option \texttt{ts\_index}, enabling users to visualize the desired time series.}

\noindent{\bf Example 5: mean change in high-dimensional time series}

We generate high-dimensional time series data based on the following simulation setting:
\begin{align*}
	(\mathrm{HD}):	Y_t= \mu_{i} + X_t, ~~ \tau_{i-1}+1\leq t\leq \tau_i, ~~ i=1,2,\cdots, 6,
\end{align*}
where $X_t$ is a sequence of i.i.d. $N(0, I_{100})$ random vectors and the five change-points are evenly located at $(\tau_1,\tau_2,\cdots,\tau_5)=(100,200,\cdots,500),$ with $\tau_0=0$ and $\tau_6=600$. We set $\mu_1=\mathbf{0}_{100}$, $\theta_i=\mu_{i+1}-\mu_i$, $\theta_i=(-1)^i (\mathbf{1}_5^\top, \mathbf{0}_{95}^\top)^\top \times \sqrt{4/5}$ for $i =1,2,\cdots,5.$ We apply \texttt{SNSeg\_HD()} to analyze this time series with \texttt{grid\_size\_scale} set at $0.05$ and \texttt{confidence} set at 0.9. 

\begin{example}
	# Generate model (HD)
	set.seed(7)
	p <- 100
	n <- 600
	cp_sets <- c(0, 100, 200, 300, 400, 500, 600)
	mean_shift <- c(0, sqrt(4 / 5), 0, sqrt(4 / 5), 0, sqrt(4 / 5))
	ts <- matrix(rnorm(n * p, 0, 1), n, p)
	no_seg <- length(cp_sets) - 1
	for (index in 1:no_seg) { # Mean shift
		tau1 <- cp_sets[index] + 1
		tau2 <- cp_sets[index + 1]
		ts[tau1:tau2, 1:5] <- ts[tau1:tau2, 1:5] + mean_shift[index]
	}
	
	# SNHD for high-dimensional means
	par(mfrow=c(2,2))
	result_hd <- SNSeg_HD(ts, confidence = 0.9, grid_size_scale = 0.05,
	plot_SN = TRUE, est_cp_loc = TRUE, 
	ts_index = c(1:4))
	
	# Output
	result_hd$est_cp
	[1]  105 203 302 397 500
\end{example}

\begin{figure}[!h]
	\centering
	\includegraphics[scale=0.5]{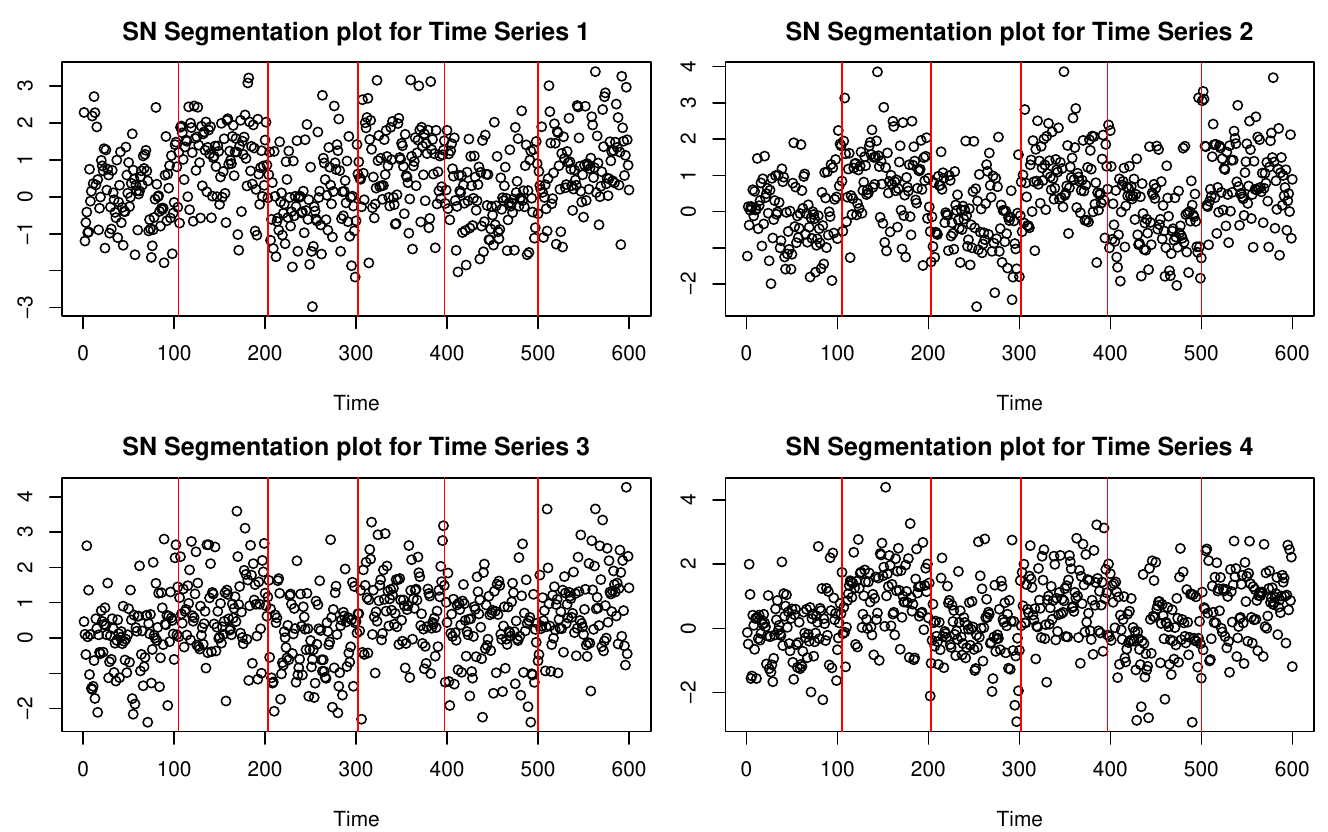}
	\caption{An example of estimating changes in high-dimensional mean for model (HD). The detected change-point locations are indicated by a red vertical line for the first 4 time series.}
	\label{fig:SN-HD-high-dimensional}
\end{figure}

{Figure \ref{fig:SN-HD-high-dimensional} plots the first four individual time series and the estimated change-points as requested by the argument \texttt{ts\_index = c(1:4)}.} As observed in the example, \texttt{SNSeg\_HD()} successfully detects all the change-points in this high-dimensional time series. This result demonstrates the effectiveness and feasibility of using SN algorithms for change-point detection in high-dimensional time series.

\subsection{Generate the SN Test Statistics }\label{subsec:teststatistic}

As discussed in Section \ref{sec:SNCP}, the success of SNCP and SNHD depends on the local SN test statistic $T_{1:n}(k)$ and $T_{1,n}^U(k)$ for $k=1,\cdots,n$. To facilitate further analysis, the function \texttt{maxSNsweep()} allows the users to compute and plot these test statistics along with the identified change-points. The \textbf{R} code is given as:
\begin{example}
	max_SNsweep(SN_result, plot_SN = TRUE, est_cp_loc = TRUE, critical_loc = TRUE)
\end{example}
It takes the following arguments:
\begin{itemize}
	\item \texttt{SN\_result}: A list generated as the output of the functions \texttt{SNSeg\_Uni()}, \texttt{SNSeg\_Multi()}, or \texttt{SNSeg\_HD()}.  
	\item \texttt{plot\_SN}: A Boolean value that specifies whether to return an SN test statistics segmentation plot.
	\item \texttt{est\_cp\_loc}: A Boolean value that specifies whether to plot a red vertical line for each estimated change-point.
	\item \texttt{critical\_loc}: A Boolean value that specifies whether to plot a blue horizontal line for the critical value $K_n$ or $K_n^U$ used in the SN test.
\end{itemize}
When called, \texttt{max\_SNsweep()} returns the maximal SN test statistic, namely $T_{1,n}(k)$ or $T_{1,n}^U(k)$, for each time point $k$. In addition, it can provide a segmentation plot based on these SN test statistics. Users are able to determine whether to mark the change-point locations and the critical value on the plot.

As an illustration of \texttt{max\_SNsweep()}, suppose we apply \texttt{SNSeg\_Uni()} to estimate change-points and save the output as \texttt{result1}. The following code can be used to generate the SN test statistic $T_{1:n}(k)$ for each $k=1,\cdots,n$ and in addition the segmentation plot, which is already given in the lower panel of Figure \ref{fig:SN-uni-single-var}.
\begin{example}
	SNstat1 <- max_SNsweep(result1, plot_SN = TRUE, est_cp_loc = TRUE, critical_loc = TRUE)
\end{example}

% \subsection{{Computation of Parameter Estimates}}\label{subsec:parameter-estimate}

{As delineated in Section \ref{subsec:SNSegUni}, \ref{subsec:SNSegMulti} and \ref{subsec:SNSegHD}, functions \texttt{SNSeg\_Uni()}, \texttt{SNSeg\_Multi()}, and \texttt{SNSeg\_HD()}  serve as the foundation for the \texttt{SNSeg\_estimate()} function, which facilitates the computation of parameter estimates for individual segments separated by the identified change-points. When called, \texttt{SNSeg\_estimate()} returns an S3 object of class \texttt{"SNSeg\_estimate"} containing the parameter estimate of each segment.  The \textbf{R} code is given as:}
\begin{example}
	SNSeg_estimate(SN_result)
\end{example}
{It takes the following argument:\begin{itemize}
		\item \texttt{SN\_result}: an S3 object with class \texttt{"SNSeg\_Uni"}, \texttt{"SNSeg\_Multi"} or \texttt{"SNSeg\_HD"}. The input of \texttt{SN\_result} must be the output from one of the functions in \texttt{SNSeg\_Uni()}, \texttt{SNSeg\_Multi()} and \texttt{SNSeg\_HD()}.
	\end{itemize}
	We refer back to \textbf{Example 1} for an illustration of its use.
}

\section{{Additional Numerical Results}}\label{sec:additional}
{This section provides additional numerical results of \pkg{SNSeg}. In Section \ref{subsec:execution-time}, we conduct a sensitivity analysis of \texttt{SNSeg\_Uni()} across various input parameters. Section \ref{subsec:comparison-of-methods} compares  with other popular change-point estimation packages. In  Section \ref{subsec:single-vs-multiple}. we demonstrate the usefulness of employing multiple parameters and contrasts it with detecting changes in a single parameter. We also provide brief explanations and recommendations on the selection of quantiles. 
	% Lastly, in Section \ref{subsec:limitations}, we enumerate some limitations of the core functions and provide recommendations regarding the choice of $\epsilon$.
}

{ In this section, we measure the accuracy of change-point estimation by counting the difference between the number of estimated change-points and true values $\hat{m}-m_o$,  the Hausdorff distance $d_H$, and adjusted Rand index (ARI). The Hausdorff distance is defined as follows. Denote the set of true change-points as $\tau_o$ and the set of estimated change-points as $\hat{\tau}$, we define $d_1(\tau_o,\hat{\tau})= \max_{\tau_1\in\hat{\tau}} \min_{\tau_2\in \tau_o} |\tau_1-\tau_2|$ and $d_2(\tau_o,\hat{\tau})= \max_{\tau_1\in\tau_o} \min_{\tau_2\in\hat{\tau}} |\tau_1-\tau_2|$, 
	where $d_1(\tau_o,\hat{\tau})$ measures the over-segmentation error of $\hat{\tau}$ and $d_2(\tau_o,\hat{\tau})$ measures the under-segmentation error of $\hat{\tau}$. The Hausdorff distance is $d_H(\tau_o, \hat{\tau})= \max \{d_1(\tau_o,\hat{\tau}),d_2(\tau_o,\hat{\tau})\}$. The ARI is originally proposed in \cite{morey1984measurement}  as a measure of similarity between two different partitions of the same observations
	for evaluating the accuracy of clustering. Under the change-point setting, we calculate the ARI between partitions of the time series given by $\hat{\tau}$ and $\tau_o$. Ranging from 0 to 1, a higher ARI indicates more coherence between the two partitions by $\hat{\tau}$ and $\tau_o$ and thus more accurate change-point estimation.
	We further note that all numerical results in this section are implemented on a laptop with 1.7 GHz 12th Gen Intel Core i7 CPU
	and 64 GB of RAM}.

\subsection{{Sensitivity analysis of SNCP}}\label{subsec:execution-time}

% {Subsequently, we scrutinize the execution time variations resulting from changes in $\epsilon$ and $K_n$, which are governed by the choice of \texttt{grid\_size\_scale} (or \texttt{grid\_size}) and the confidence level. The comparison involves $\epsilon$ values of 0.05, 0.1, 0.15, 0.2, and 0.3, alongside confidence levels of 0.9, 0.95, 0.99, 0.995, and 0.999. The execution times for \texttt{SBSeg\_Uni()} are compiled and averaged over 100 replications for model (V1), where the time series is examined based on the variation in a single mean parameter. Table \ref{table:execution-time-epsilon} presents the tabulated results.}

{We first examine the performance variations resulted from choices of the trimming parameter $\epsilon$ and threshold $K_n$ (reflected by the confidence level $q$) when multiple change-points exist. Specifically, we generate the data according to model (SA):
	% $$
	% (\mathrm{SA} 1): n=1200, \rho=0,  Y_t= \begin{cases}
		% 0+X_t, & t \in  [1, 150], [301, 450], [601, 750], [901, 1050]
		% \\ \delta+X_t, & t \in  [151, 300], [451, 600], [751, 900], [1051, 1200] .
		% \end{cases} 
	% $$
	$$
	(\mathrm{SA}): n=1200, \rho=0.5,  Y_t= \begin{cases}
		0+X_t, & t \in  [1, 150], [301, 450], [601, 750], [901, 1050]
		\\ \delta+X_t, & t \in  [151, 300], [451, 600], [751, 900], [1051, 1200],
	\end{cases}
	$$
	where $\{X_t\}_{t=1}^n$ is generated from a unit-variance  AR(1) process with $X_t=  X_{t-1}/2+\sqrt{3}\epsilon_t/2$, and $\{\epsilon_t\}$ is \textit{i.i.d.} $N(0,1)$. We vary $\delta\in\{\sqrt{3},\sqrt{6}\}$ to compare the results under low and high signal-to-noise ratios.  } 
% \textcolor{red}{Are you re-running the experiments on your own PC? But later you state that the Table is copied from Zhao et al. I suggest we do not mention Table 1 is copied from Zhao.}

\begin{table}[!h]
	\begin{center}
		\caption{Sensitivity analysis of $\epsilon$ and $q$ for time series model (SA) with $\delta\in\{\sqrt{3},\sqrt{6}\}$.} \label{table:sensitivity-analysis}
		\scalebox{0.9}{\begin{tabular}{c|c|ccccccc|c|c|c|c|c}
				\hline \hline
				&  & &  & & $\hat{m}-m_o$ & & & \\ \hline
				$\delta$ & $(q,\epsilon)$ & $\leq -3$ & $-2$ & $-1$ & $0$ & $1$ & $2$ & $\geq 3$ & ARI & $d_1$ &  $d_2$ & $d_H$ & time \\ \hline
				\multirow{10}{*}{$\sqrt{3}$} & 0.90, 0.05 & 0 & 1 & 89 & 882 & 27 & 1 & 0 & 0.933 & 1.12 & 2.18 & 2.18 & 1.59 \\& 0.95, 0.05 & 0 & 26 & 175 & 788 & 11 & 0 & 0 & 0.918 & 1.01 & 3.38 & 3.38 & 1.59 \\ & 0.90, 0.08 & 0 & 33 & 195 & 772 & 0 & 0 & 0 & 0.911 & 0.97 & 3.65 & 3.65 & 0.59 \\ & 0.95, 0.08 & 16 & 85 & 302 & 597 & 0 & 0 & 0 & 0.880 & 0.92 & 5.92 & 5.92 & 0.59\\ & 0.90, 0.10 & 0 & 4 & 58 & 938 & 0 & 0 & 0 & 0.931 & 1.02 & 1.75 & 1.75 & 0.36 \\ & 0.95, 0.10 & 0 & 24 & 120 & 856 & 0 & 0 & 0 & 0.919 & 0.99 & 2.74 & 2.74 & 0.36 \\ & 0.90, 0.12 & 0 & 4 & 130 & 866 & 0 & 0 & 0 & 0.933 & 0.77 & 2.17 & 2.17 & 0.24 \\ & 0.95, 0.12 & 1 & 14 & 159 & 826 & 0 & 0 & 0 & 0.926 & 0.76 & 2.69 & 2.69 & 0.24 \\ & 0.90, 0.15 & 1000 & 0 & 0 & 0 & 0 & 0 & 0 & 0.002 & 0.00 & 49.98 & 49.98 & 0.13 \\ & 0.95, 0.15 & 1000 & 0 & 0 & 0 & 0 & 0 & 0 & 0.001 & 0.00 & 50.01 & 50.01 & 0.13 \\
				\hline
				\multirow{10}{*}{$\sqrt{6}$} & 0.90, 0.05 & 0 & 0 & 4 & 964 & 31 & 1 & 0 & 0.965 & 0.77 & 0.60 & 0.77 & 1.60 \\& 0.95, 0.05 & 0 & 0 & 11 & 968 & 21 & 0 & 0 & 0.965 & 0.66 & 0.79 & 0.79 & 1.60 \\ & 0.90, 0.08 & 0 & 0 & 16 & 984 & 0 & 0 & 0 & 0.963 & 0.58 & 0.77 & 0.77 & 0.67 \\ & 0.95, 0.08 & 0 & 2 & 48 & 950 & 0 & 0 & 0 & 0.958 & 0.57 & 1.17 & 1.17 & 0.67 \\ & 0.90, 0.10 & 0 & 0 & 2 & 998 & 0 & 0 & 0 & 0.961 & 0.62 & 0.65 & 0.65 & 0.37 \\ & 0.95, 0.10 & 0 & 0 & 8 & 992 & 0 & 0 & 0 & 0.961 & 0.62 & 0.72 & 0.72 & 0.37 \\ & 0.90, 0.12 & 0 & 0 & 34 & 966 & 0 & 0 & 0 & 0.958 & 0.59 & 0.95 & 0.95 & 0.25 \\ & 0.95, 0.12 & 0 & 0 & 34 & 966 & 0 & 0 & 0 & 0.958 & 0.59 & 0.95 & 0.95 & 0.25 \\ & 0.90, 0.15 & 1000 & 0 & 0 & 0 & 0 & 0 & 0 & 0.000 & 0.00 & 50.01 & 50.01 & 0.16 \\ & 0.95, 0.15 & 1000 & 0 & 0 & 0 & 0 & 0 & 0 & 0.000 & 0.00 & 50.01 & 50.01 & 0.16 \\
				\hline
				\hline   
		\end{tabular}}
	\end{center}  
\end{table}
{ We vary $\epsilon \in \{0.05, 0.08, 0.10, 0.12, 0.15\}$ and $q\in \{0.90, 0.95\}$, and study how they affect the performance of SNCP. The numerical result over 1000 replications is summarized in Table \ref{table:sensitivity-analysis} for reader's convenience.  From the table, we find that as long as the window size  $\epsilon$ is smaller than the minimum spacing $\epsilon_o=0.125$, the performance of SNCP is quite robust and stable across the choices of $\epsilon$ and $q$. However,  SNCP fails to detect changes with the window size $\epsilon=0.15>\epsilon_o$, highlighting the importance of selecting an appropriate value for $\epsilon$. Furthermore, we find that the execution time increases with diminishing values of $\epsilon$, while no discernible disparity in execution time is found across various thresholds.
}
% \begin{table}
	%   \begin{center}
		%   \caption{Execution time of SNCP for different $\epsilon$ and confidence levels in seconds.} \label{table:execution-time-epsilon}
		% \begin{tabular}{ccccccccc}
			% $\epsilon$(\texttt{grid\_size\_scale}) & Time & Confidence Level & Time \\
			%   \hline   
			% 0.05 & 1.7743 & 0.9 & 0.0256 \\
			% 0.1 & 0.3911 & 0.95 & 0.0264 \\
			% 0.15 & 0.1544 & 0.99 & 0.0241 \\
			% 0.2 & 0.0722 & 0.995 & 0.0243 \\
			% 0.3 & 0.0245 & 0.999 & 0.0248 \\
			%     \hline
			% \end{tabular}
		%   \end{center}  
	% \end{table}

{ We also briefly study the execution time of  SNCP using \texttt{SNSeg\_Uni()} across multiple model parameters including mean, variance, autocorrelation (ACF), 90\% quantile, and a multi-parameter scenario with both variance and  90\% quantile.  For model (V1) from \textbf{Example 1} and (MP1) from \textbf{Example 3}, Table \ref{table:execution-time-parameter} presents the averaged execution time over 100 replications of \texttt{SBSeg\_Uni()}. The execution time of SNCP varies in the order of mean, variance, ACF, and quantile, progressing from the lowest to the highest. Notably, the multi-parameter scenario requires a longer runtime compared to the single-parameter cases.}
\begin{table}[!h]
	\begin{center}
		\caption{Execution time (in seconds) of SNCP for different parameters when applied to the models (V1) and (MP1) averaged over 100 replications.}\label{table:execution-time-parameter}
		\scalebox{0.9}{
			\begin{tabular}{ccccccccc}
				\hline\hline
				Model & mean & variance & ACF & quantile & multi-parameter\\
				\hline
				V1 & 1.75 & 5.86 & 8.83 & 17.80 & 31.83 \\
				\hline
				MP1 & 1.68 & 5.82 & 8.52 & 16.56 & 31.06 \\
				\hline\hline
		\end{tabular}}
	\end{center}  
\end{table}

\subsection{{Comparison: SNCP vs BinSeg, PELT, MOSUM and ECP}}\label{subsec:comparison-of-methods}
{We next compare SNCP with BinSeg, PELT, MOSUM and ECP in terms of the accuracy of change-point estimates, especially when data exhibits temporal dependence. BinSeg and PELT are implemented by  the package \textbf{changepoint} (the functions \texttt{cpt.mean()} and \texttt{cpt.var()}, respectively), MOSUM  utilizes the package \textbf{mosum} (the function \texttt{mosum()}), and ECP adopts the package \textbf{ecp} (the function \texttt{e.divisive()}). For SNCP, we set the trimming parameter $\epsilon=0.05$ and confidence level $q=0.9$, adhering to its default configuration. The input parameters for other methods are also chosen as default values.  In particular, the default thresholds for MOSUM and ECP are based on  critical values of asymptotic null distribuions under confidence level 0.95; while that for BinSeg and PELT are  non-asymptotic.  For the bandwidth parameter \texttt{G} that requires manual selection in the function \texttt{mosum()} for MOSUM,   we  let $\texttt{G=100}$. This choice aligns with the recommendation in Section 3.5 from \cite{eichinger2018mosum} that \texttt{G} should be half the minimal distance between two change points. In the case of model (M) below, this minimum distance is 200. 
}

{We first compare the performance of all methods under the no change-point scenario, where the time series is stationary with no change-point. We simulate a stationary univariate time series $\{Y_t\}_{t=1}^{n=1000}$ from a unit-variance AR(1) process $Y_t=\rho Y_{t-1}+\sqrt{1-\rho^2}\epsilon_t$ where $\{\epsilon_t\}$ is \textit{i.i.d.}  $N(0,1)$. We vary $\rho\in\{0,0.4,0.7\}$ to investigate the robustness of SNCP (and other methods) against false positives (i.e. type-I error) under different levels of temporal dependence. }

{The experiment is repeated 1000 times for each $\rho$, and the results are documented in Table \ref{table:comparison-no-change-point}.   
	In general, under their respective default settings, all methods provide satisfactory type-I error control when there is no temporal dependence ($\rho=0$)  whereas MOSUM and ECP are prone to produce false positives when dependence is moderate ($\rho=0.4$). Under strong temporal dependence ($\rho=0.7$), all tests exhibit high false-positive rates and SNCP is the most robust option. }

% \begin{table}[!ht]
	%   \begin{center}
		%   \caption{Number of change-points detected of each method  when there is no change-point.} \label{table:comparison-no-change-point}
		% \scalebox{0.9}{\begin{tabular}{c|ccc|ccc|ccc}
				% \hline \hline
				% $n=600$ &  & $\rho=0$  &  &  & $\rho=0.4$ &  & & $\rho=0.7$ &  \\ \hline
				% $\hat{m}$ & $0$ & $1$ & $\geq 2$ & $0$ & $1$ & $\geq 2$ & $0$ & $1$ & $\geq2$  \\ \hline
				% SNCP & 925 & 67 & 8 & 852 & 132  & 16 & 628 & 280 & 92 \\ \hline
				% BinSeg & 999 & 1 & 0 & 950 & 49 & 1 & 582 & 213 & 205 \\ \hline
				% PELT & 1000 & 0 & 0 & 933 & 44 & 23 & 234 & 91 & 675 \\ \hline
				% MOSUM & 956 & 38 & 6 & 426 & 324 & 250 & 19 & 94 & 887 \\ \hline
				% ECP & 941 & 29 & 30 & 237 & 112 & 651 & 0 & 1 & 999 \\ \hline
				%   \hline 
				% \end{tabular}}
		%   \end{center}  
	% \end{table}

\begin{table}[!ht]
	\begin{center}
		\caption{Number of change-points detected of each method  when there is no change-point.} \label{table:comparison-no-change-point}
		\scalebox{0.9}{\begin{tabular}{c|ccc|ccc|ccc}
				\hline \hline
				$n=1000$ &  & $\rho=0$  &  &  & $\rho=0.4$ &  & & $\rho=0.7$ &  \\ \hline
				$\hat{m}$ & $0$ & $1$ & $\geq 2$ & $0$ & $1$ & $\geq 2$ & $0$ & $1$ & $\geq2$  \\ \hline
				SNCP & 910 & 82 & 8 & 884 & 111  & 5 & 744 & 210 & 46 \\ \hline
				BinSeg & 1000 & 0 & 0 & 963 & 35 & 2 & 558 & 240 & 202 \\ \hline
				PELT & 1000 & 0 & 0 & 943 & 30 & 27 & 152 & 66 & 782 \\ \hline
				MOSUM & 954 & 43 & 3 & 292 & 330 & 378 & 7 & 16 & 977 \\ \hline
				ECP & 952 & 24 & 24 & 145 & 72 & 783 & 0 & 0 & 1000 \\ \hline
				\hline 
		\end{tabular}}
	\end{center}  
\end{table}

{
	We further examine their power performance under model (M):
	$$
	(\mathrm{M}): n=1000, \rho\in\{0,0.4,0.7\},  Y_t= \begin{cases}
		X_t, & t \in[1,200], [401,600], [801,1000]\\ 2+X_t, & t \in[201,400], [601,800] .
	\end{cases}
	$$
	% $$
	% (\mathrm{M} 3): n=500, \rho=-0.7,  Y_t= \begin{cases}
		% 0.4+X_t, & t \in[1,250], [376,500]\\ X_t, & t \in[251,375] .
		% \end{cases}
	% $$
	Here  $\{X_t\}_{t=1}^{n}$ is generated from a unit-variance AR(1) process that $X_t=\rho X_{t-1}+\sqrt{1-\rho^2}\epsilon_t$, and $\{\epsilon_t\}$ is \textit{i.i.d.} $N(0,1)$. The true change-points occur at 200, 400, 600 and 800.  Table \ref{table:comparison-of-methods} summarizes the results over 1000 replications.

	From the table, we observe that in the absence of temporal dependence ($\rho=0$), all the methods perform well, with PELT being the most effective. It should be noted that, due to the use of self-normalizer, SNCP may experience some decrease in estimation accuracy compared to other methods.  When the dependence is moderate at $\rho=0.4$, SNCP demonstrates robust performance, while other competing methods tend to overestimate the number of change-points. With stronger dependence ($\rho=0.7$), SNCP exhibits the best performance based on the distribution of $\hat{m}-m_o$ along with $d_H$ and ARI, while all the other methods severely overestimate the number of change-points.  In terms of the computational speed, we find that BinSeg, PELT, and MOSUM are more efficient than SNCP. Consequently, we recommend employing SNCP for change-point estimation when data exhibit moderate or strong dependence, while opting for BinSeg or PELT in cases with no or weak  dependence.} 

{Here, we only compare the results under mean shifts, and we refer the interested readers to \cite{zhao2021segmenting} for results in other settings. Broadly speaking, our findings indicate that the SNCP exhibits greater robustness to temporal dependence compared to competing methods. However, it's worth noting that other methods might demonstrate superior performance in instances where temporal dependence is weak. For instance, BinSeg, PELT, and MOSUM are adept at handling frequent change-points with fast computational speed.}

\begin{table}[!h]
	\begin{center}
		\caption{Performance of different methods for the time series model (M).} \label{table:comparison-of-methods}
		\scalebox{0.85}{\begin{tabular}{c|c|ccccccc|c|c|c|c|c}
				\hline \hline
				&  & &  & & $\hat{m}-m_o$ & & & \\ \hline
				$\rho$ & Method & $\leq -3$ & $-2$ & $-1$ & $0$ & $1$ & $2$ & $\geq 3$ & ARI & $d_1$ &  $d_2$ & $d_H$ & time \\ \hline
				\multirow{5}{*}{$\rho=0$} & SNCP & 0 & 0 & 0 & 991 & 9 & 2 & 0 & 0.983 & 4.13 & 3.42 & 4.13 & 6.85 \\& BinSeg & 0 & 0 & 0 & 961 & 39 & 0 & 0 & 0.988 & 3.83 & 3.26 & 3.83& 0.01 \\ & PELT & 0 & 0 & 0 & 999 & 1 & 0 & 0 & 0.994 & 1.73 & 1.69 & 1.73 & 0.11 \\ & MOSUM & 0 & 0 & 0 & 993 & 7 & 0 & 0 & 0.992 & 3.09 & 2.09 & 3.09 & 0.00\\ & ECP & 0 & 0 & 0 & 937 & 56 & 7 & 0 & 0.984 & 7.04 & 2.24 & 7.04 & 46.85 \\ \hline
				\multirow{5}{*}{$\rho=0.4$} & SNCP & 0 & 0 & 0 & 972 & 27 & 1 & 0 & 0.956 & 8.10 & 5.73 & 8.10 & 5.78 \\& BinSeg & 0 & 0 & 0 & 744 & 256 & 0 & 0 & 0.963 & 15.16 & 5.60 & 15.16 & 0.00 \\ & PELT & 0 & 0 & 0 & 862 & 109 & 29 & 0 & 0.964 & 14.03 & 3.81 & 14.03 & 0.05 \\ & MOSUM & 0 & 0 & 0 & 779 & 199 & 20 & 2 & 0.959 & 38.19 & 4.81 & 38.19 & 0.00\\ & ECP & 0 & 0 & 0 & 170 & 184 & 217 & 429 & 0.828 & 87.05 & 4.14 & 87.05 & 54.62 \\ \hline
				\multirow{5}{*}{$\rho=0.7$} & SNCP & 0 & 2 & 59 & 865 & 69 & 4 & 1 & 0.934 & 17.83 & 23.69 & 29.74 & 6.01 \\& BinSeg & 0 & 0 & 0 & 201 & 799 & 0 & 0 & 0.930 & 55.11 & 11.30 & 55.11 & 0.00 \\ & PELT & 0 & 0 & 0 & 104 & 162 & 195 & 539 & 0.867 & 99.50 & 8.95 & 99.50 & 0.06 \\ & MOSUM & 0 & 0 & 0 & 209 & 414 & 281 & 96 & 0.909 & 126.3 & 11.06 & 126.3 & 0.00\\ & ECP & 0 & 0 & 0 & 1 & 0 & 1 & 998 & 0.537 & 146.4 & 8.31 & 146.4 & 92.2 \\ \hline
				\hline   
				
		\end{tabular}}
	\end{center}  
\end{table}

\subsection{{Single vs Multiple Parameters}}\label{subsec:single-vs-multiple}

{As outlined in Section \ref{subsec:SNSegUni}, the function \texttt{SNSeg\_Uni()} enables users to examine changes in either a single parameter or multiple parameters. We first provide a simple example which shows that using multiple parameters may not necessarily be significantly inferior to using a single parameter, in terms of change-point estimates.  This observation holds true even when the change solely stems from the single parameter. For example, we consider model (M) with $\rho=0.4$ from Section \ref{subsec:comparison-of-methods}, which is solely driven by mean changes. }

% $$
% \begin{split}
	%     (\mathrm{M} 2)&: n=500, \rho=-0.7, Y_t= \begin{cases}
		% -3+X_t, & t \in[1,250], [376,500]\\ 1+X_t, & t \in [251,375].
		% \end{cases} \\
	%     (\mathrm{M} 3)&: n=600, \rho=0.2, Y_t= \begin{cases}
		% X_t, & t \in[1,100], [201,300], [401,500], \\ 2+X_t, & t \in [101,200], [301,400], [501,600].
		%     \end{cases}
	% \end{split}
% $$

% where $X_t$ is simulated from an AR(1) process with $X_t=\rho X_{t-1}+\epsilon_t$ and $\{\epsilon_t\}$ is \textit{i.i.d.} $N(0,1)$. As outlined in the scenario (M2), $Y_t$ is only characterized by the variation in mean with the actual change-points occuring at 250 and 375. For (M3) the actual change-points occur at 100, 200, 300, 400 and 500. Using the function \texttt{SNSeg\_Uni()}, we compare the single mean change with a composite change involving both the mean and the variance in (M2). For (M3) over 1000 replications. 

\begin{table}[!h]
	\begin{center}
		\caption{Performance of SNCP based on a single parameter and multiple parameters for (M). Beginning with "SN", $\text{M, V, Q}_{20}$ represent mean, variance and 20\% quantile, respectively.} \label{table:comparison-of-multiple-parameters}
		\scalebox{0.85}{\begin{tabular}{c|c|ccccccc|c|c|c|c|c}
				\hline \hline
				&  & &  & & $\hat{m}-m_o$ & & & \\ \hline
				Model & Method & $\leq -3$ & $-2$ & $-1$ & $0$ & $1$ & $2$ & $\geq 3$ & ARI & $d_1$ &  $d_2$ & $d_H$ & time \\ \hline
				\multirow{3}{*}{(M)} & SNM & 0 & 0 & 0 & 972 & 28 & 0 & 0 & 0.969 & 8.54 & 6.40 & 8.54 & 1.49 \\& $\text{SNMQ}_{20}$ & 0 & 0 & 3 & 958 & 39 & 0 & 0 & 0.967 & 9.71 & 7.33 & 10.25 & 28.6 \\ & SNMV & 0 & 0 & 3 & 934 & 62 & 1 & 0 & 0.963 & 12.53 & 8.06 & 13.11 & 18.2 \\ \hline
				\hline  
				% \multirow{2}{*}{(M2)} & Mean & 0 & 0 & 0 & 1000 & 0 & 0 & 0 & 0.996 & 0.590 & 0.590 & 0.590 & 2.29 \\& Multiple1 & 0 & 0 & 0 & 957 & 40 & 3 & 0 & 0.987 & 5.167 & 0.578 & 5.167 & 38.3 \\ \hline
				% \multirow{2}{*}{(M3)} & Mean & 0 & 0 & 7 & 975 & 17 & 1 & 0 & 0.962 & 5.027 & 5.016 & 5.701 & 2.70 \\& Multiple2 & 0 & 0 & 24 & 960 & 16 & 0 & 0 & 0.959 & 4.944 & 6.678 & 7.238 & 128.8 \\ \hline
		\end{tabular}}
	\end{center}  
\end{table}

% { Table \ref{table:comparison-of-multiple-parameters} summarizes the numerical results over 1000 replications. As indicated by the result, the combination of mean and variance achieves more satisfactory performance. This can be illustrated by the distribution of $\hat{m}-m_o$, a higher value of ARI and a smaller value of Hausdorff distance $d_H$ for the multiple-parameters case. However, it is crucial to acknowledge that using multiple parameters may not always be superior to applying a single parameter for univariate time series analysis.}

{ Table \ref{table:comparison-of-multiple-parameters} summarizes the numerical results over 1000 replications. For clarity, we specify these cases using names beginning with "SN". For instance, $\text{SNM}$ denotes SNCP for estimating changes in a single mean, $\text{SNMQ}_{20}$ represents SNCP for estimating changes in both mean and the 20\% quantile, and $\text{SNMV}$ targets the mean and variance changes simultaneously. From the table, we find that all three methods yield rather similar results, albeit mild overestimation by SNMV. This indicates that introducing additional parameters does not necessarily hinder the performance of SNCP.}

{We then provide another example that examining multiple parameters can outperform examining a single parameter. Specifically, we compare the performance of SNCP based on a single variance or quantile (90\% or 95\%) and their multi-parameter combination under the setting (MP1) from \textbf{Example 3}. Recall that for (MP1), the change originates from the upper quantiles and the actual change-points take place at 333 and 667. The numerical result of (MP1) over 1000 replications is taken from Table 5 of \cite{zhao2021segmenting}, and summarized in Table \ref{table:single-vs-multiple-mp1} here for readers' convenience. Similar to  Table \ref{table:comparison-of-multiple-parameters}, we specify the parameter settings using names beginning with "SN". For instance, $\text{SNV}$ denotes the change in a single variance, $\text{SNQ}_{90}$ represents the change in the 90\% quantile, and $\text{SNQ}_{90}\text{V}$ targets the variance and 90\% quantile changes simultaneously. We observe that for (MP1), $\text{SNQ}_{90}$ and $\text{SNQ}_{95}$ performs well with a high estimation accuracy since the change of (MP1) originates from upper quantiles. By integrating changes in variance and quantiles, improvements are observed across estimation accuracy, ARI, and Hausdorff distance $d_H$. Notably, the combined-parameter setting $\text{SNQ}_{90,95}\text{V}$ achieves the optimal performance compared to all the other parameter configurations. }

\begin{table}[!h]
	\begin{center}
		\caption{Performance of SNCP based on the change in a single parameter and multiple parameters for (MP1). $Q_{90},Q_{95},V$ represent the change in the 90th and the 95th quantile as well as the variance respectively.}\label{table:single-vs-multiple-mp1}
		\scalebox{0.8}{\begin{tabular}{c|c|ccccccc|c|c|c|c}
				\hline \hline
				&  & &  & & $\hat{m}-m_o$ & & & \\ \hline
				Model & Method & $\leq -3$ & $-2$ & $-1$ & $0$ & $1$ & $2$ & $\geq 3$ & ARI & $d_1$ &  $d_2$ & $d_H$ \\ \hline
				\multirow{7}{*}{(MP1)} & $\text{SNQ}_{90}$ & 0 & 10 & 132 & 805 & 50 & 3 & 0 & 0.839 & 3.25 & 7.26 & 7.85  \\ & $\text{SNQ}_{95}$ & 0 & 5 & 100 & 820 & 73 & 2 & 0 & 0.868 & 3.16 & 5.70 & 6.62  \\ & $\text{SNV}$ & 0 & 2 & 110 & 832 & 54 & 2 & 0 & 0.869 & 2.45 & 5.47 & 6.06  \\ & $\text{SNQ}_{90,95}$ & 0 & 3 & 82 & 850 & 62 & 3 & 0 & 0.878 & 3.01 & 4.88 & 5.67  \\ & $\text{SNQ}_{90}\text{V}$ & 0 & 0 & 56 & 869 & 70 & 5 & 0 & 0.891 & 3.04 & 3.95 & 4.77  \\ & $\text{SNQ}_{95}\text{V}$ & 0 & 2 & 64 & 861 & 68 & 5 & 0 & 0.889 & 2.92 & 4.30 & 5.14  \\ & $\text{SNQ}_{90,95}\text{V}$ & 0 & 2 & 48 & 882 & 66 & 2 & 0 & 0.894 & 2.95 & 3.79 & 4.58 \\ \hline 
				\hline   
		\end{tabular}}
	\end{center}  
\end{table}

{Overall, our findings illustrate that employing multiple parameters does not always diminish performance of SNCP compared to using a single parameter, even when the change is primarily driven by a single parameter. Nevertheless, it is important to recognize that incorporating prior information on change types can enhance the effectiveness of SNCP. }

% \textcolor{purple}{Shubo: I summarized Section 4.4 into 2 paragraphs shown below. I didn't remove section 4.4, in the case that someone wants to take a look at it. Also, I think we need to rename Section 4.3 because the choice of quantiles is now incorporated.}

{Another aspect that is of interest is the choice of quantile for SNCP. As delineated in Section \ref{subsec:SNSegUni}, the \texttt{SNSeg\_Uni()} function provides users with the capability to assess variations in either a single or multiple quantiles. Particularly for practitioners, an appropriate selection of the quantile becomes pivotal when the true quantile that may change remains unknown. Table \ref{table:single-vs-multiple-mp1} for model (MP1) also offers valuable insights in this regard. Given that (MP1) experiences changes in upper quantiles, the usage of the 90\% or the 95\% quantile yields satisfactory results. Furthermore, the application of both 90\% and 95\% quantiles in combination results in an improvement compared to utilizing a single quantile.}

{Consequently, in cases where the specific quantile that is changing is unknown, we recommend users visually inspect their time series for signals such as peaks or troughs to assess the potential range of quantiles where changes might occur, and further employ multiple quantiles within this range for more robust change-point estimation. In other words, if one knows that the change happens in a specific range of the distribution (for example, the upper tail), we recommend he/she target several quantiles in this range (for example, targeting 90\%, 95\%) simultaneously, instead of picking only one quantile. In practice, it is seldom that only a particular quantile changes, while the other quantile levels near this quantile exhibit no change. Hence, testing several quantiles together can boost power and estimation accuracy to the best degree.}

\section{Summary}\label{sec:conclusion}
In this paper, we introduced the  \textbf{R} package \pkg{SNSeg}, which provides implementations of the SN-based procedures for change-points estimation in univariate, multivariate, and high-dimensional time series. We described the main functions of the package, namely \texttt{SNSeg\_Uni()}, \texttt{SNSeg\_Multi()}, \texttt{SNSeg\_HD()}, which enable the detection of change-points in a single or multiple parameter(s) of the time series. Furthermore, we presented examples demonstrating the usage of the package, including visualizing both the time series data and the segmentation plots of the SN test statistics, {as well as the computation of parameter estimates within the segments that are separated by the estimated change-points}.

The \pkg{SNSeg} package offers a comprehensive set of tools to effectively identify change-points in time series data. We hope the availability of \pkg{SNSeg} on CRAN can help facilitate the analysis and understanding of temporal patterns and dynamics for both researchers and practitioners.

\bibliography{your-article.bib}

\begin{thebibliography}{21}
\providecommand{\natexlab}[1]{#1}
\providecommand{\url}[1]{\texttt{#1}}
\expandafter\ifx\csname urlstyle\endcsname\relax
  \providecommand{\doi}[1]{doi: #1}\else
  \providecommand{\doi}{doi: \begingroup \urlstyle{rm}\Url}\fi

\bibitem[Andrews(1991)]{andrews1991heteroskedasticity}
D.~W. Andrews.
\newblock Heteroskedasticity and autocorrelation consistent covariance matrix
  estimation.
\newblock \emph{Econometrica}, pages 817--858, 1991.

\bibitem[Auger and Lawrence(1989)]{auger1989algorithms}
I.~E. Auger and C.~E. Lawrence.
\newblock Algorithms for the optimal identification of segment neighborhoods.
\newblock \emph{Bulletin of mathematical biology}, 51\penalty0 (1):\penalty0
  39--54, 1989.

\bibitem[Eichinger and Kirch(2018)]{eichinger2018mosum}
B.~Eichinger and C.~Kirch.
\newblock A mosum procedure for the estimation of multiple random change
  points.
\newblock \emph{Bernoulli}, 24\penalty0 (1):\penalty0 526--564, 2018.

\bibitem[James and Matteson(2014)]{ecp2014}
N.~A. James and D.~S. Matteson.
\newblock {ecp}: An {R} package for nonparametric multiple change point
  analysis of multivariate data.
\newblock \emph{Journal of Statistical Software}, 62\penalty0 (7):\penalty0
  1--25, 2014.

\bibitem[Killick and Eckley(2014)]{changepoint}
R.~Killick and I.~A. Eckley.
\newblock {changepoint}: An {R} package for changepoint analysis.
\newblock \emph{Journal of Statistical Software}, 58\penalty0 (3):\penalty0
  1--19, 2014.

\bibitem[Killick et~al.(2012)Killick, Fearnhead, and Eckley]{Killick2012}
R.~Killick, P.~Fearnhead, and I.~A. Eckley.
\newblock Optimal detection of changepoints with a linear computational cost.
\newblock \emph{Journal of the American Statistical Association}, 107\penalty0
  (500):\penalty0 1590--1598, 2012.

\bibitem[Meier et~al.(2021)Meier, Kirch, and Cho]{mosum2021}
A.~Meier, C.~Kirch, and H.~Cho.
\newblock {mosum}: A package for moving sums in change-point analysis.
\newblock \emph{Journal of Statistical Software}, 97\penalty0 (8):\penalty0
  1--42, 2021.

\bibitem[Morey and Agresti(1984)]{morey1984measurement}
L.~C. Morey and A.~Agresti.
\newblock The measurement of classification agreement: An adjustment to the
  rand statistic for chance agreement.
\newblock \emph{Educational and Psychological Measurement}, 44\penalty0
  (1):\penalty0 33--37, 1984.

\bibitem[Newey and West(1987)]{newey1986simple}
W.~K. Newey and K.~D. West.
\newblock A simple, positive semi-definite, heteroskedasticity and
  autocorrelationconsistent covariance matrix.
\newblock \emph{Econometrica}, 55:\penalty0 703--708, 1987.

\bibitem[Ross(2015)]{cpm2015}
G.~J. Ross.
\newblock Parametric and nonparametric sequential change detection in {R}: The
  {cpm} package.
\newblock \emph{Journal of Statistical Software}, 66\penalty0 (3):\penalty0
  1--20, 2015.

\bibitem[Shao(2010)]{shao2010self}
X.~Shao.
\newblock A self-normalized approach to confidence interval construction in
  time series.
\newblock \emph{Journal of the Royal Statistical Society: Series B},
  72\penalty0 (3):\penalty0 343--366, 2010.

\bibitem[Shao(2015)]{shao2015}
X.~Shao.
\newblock Self-normalization for time series: a review of recent developments.
\newblock \emph{Journal of the American Statistical Association}, 110:\penalty0
  1797--1817, 2015.

\bibitem[Shao and Zhang(2010)]{shaozhang2010testing}
X.~Shao and X.~Zhang.
\newblock Testing for change points in time series.
\newblock \emph{Journal of the American Statistical Association}, 105\penalty0
  (491):\penalty0 1228--1240, 2010.

\bibitem[Sun et~al.(2023)Sun, Zhao, Jiang, and Shao]{SNSeg2023}
S.~Sun, Z.~Zhao, F.~Jiang, and X.~Shao.
\newblock \emph{SNSeg: Self-Normalization(SN) Based Change-Point Estimation for
  Time Series}, 2023.
\newblock URL \url{https://CRAN.R-project.org/package=SNSeg}.
\newblock R package version 1.0.0.

\bibitem[Wang and Zou(2023)]{wang2022cpss}
G.~Wang and C.~Zou.
\newblock cpss: an package for change-point detection by sample-splitting
  methods.
\newblock \emph{Journal of Quality Technology}, 55:\penalty0 61--74, 2023.

\bibitem[Wang et~al.(2022)Wang, Zhu, Volgushev, and Shao]{wang2022inference}
R.~Wang, C.~Zhu, S.~Volgushev, and X.~Shao.
\newblock Inference for change points in high-dimensional data via
  selfnormalization.
\newblock \emph{The Annals of Statistics}, 50\penalty0 (2):\penalty0 781--806,
  2022.

\bibitem[Zeileis et~al.(2002)Zeileis, Leisch, Hornik, and
  Kleiber]{zeileis2002struc}
A.~Zeileis, F.~Leisch, K.~Hornik, and C.~Kleiber.
\newblock strucchange: An r package for testing for structural change in linear
  regression models.
\newblock \emph{Journal of Statistical Software}, 7\penalty0 (2):\penalty0
  1--38, 2002.

\bibitem[Zeileis et~al.(2003)Zeileis, Kleiber, Kr\"amer, and
  Hornik]{zeileis2003cmpstat}
A.~Zeileis, C.~Kleiber, W.~Kr\"amer, and K.~Hornik.
\newblock Testing and dating of structural changes in practice.
\newblock \emph{Computational Statistics \& Data Analysis}, 44\penalty0
  (1--2):\penalty0 109--123, 2003.

\bibitem[Zhao et~al.(2021)Zhao, Jiang, and Shao]{zhao2021v1}
Z.~Zhao, F.~Jiang, and X.~Shao.
\newblock Segmenting time series via self-normalization.
\newblock \emph{arXiv preprint https://arxiv.org/pdf/2112.05331v1.pdf}, 2021.

\bibitem[Zhao et~al.(2022)Zhao, Jiang, and Shao]{zhao2021segmenting}
Z.~Zhao, F.~Jiang, and X.~Shao.
\newblock Segmenting time series via self-normalisation.
\newblock \emph{Journal of the Royal Statistical Society Series B: Statistical
  Methodology}, 84\penalty0 (5):\penalty0 1699--1725, 2022.

\bibitem[Zou et~al.(2020)Zou, Wang, and Li]{zou2020consistent}
C.~Zou, G.~Wang, and R.~Li.
\newblock Consistent selection of the number of change-points via
  sample-splitting.
\newblock \emph{Annals of Statistics}, 48\penalty0 (1):\penalty0 413, 2020.

\end{thebibliography}

\address{Shubo Sun\\
	University of Miami\\
	Herbert Business School\\
	Coral Gables, FL, USA\\
	\email{sxs3935@miami.edu}}

\address{Zifeng Zhao\\
	University of Notre Dame\\
	Mendoza College of Business\\
	Notre Dame, IN, USA\\
	\email{zifeng.zhao@nd.edu}}

\address{Feiyu Jiang\\
	Fudan University\\
	School of Management\\
	Shanghai, China\\
	\email{jiangfy@fudan.edu.cn}}

\address{Xiaofeng Shao\\
	University of Illinois at Urbana-Champaign\\
	Department of Statistics\\
	Champaign, IL, USA\\
	\email{xshao@illinois.edu}}
\end{article}

\end{document}